%% file: main.tex
\documentclass[camera,letterpaper,nomarginnotes,nonarrowgutter]{jpaper}

\input{packages}
\input{commands}

\newcommand{\thetitle}{CIPHERMATCH: Accelerating Homomorphic Encryption-Based String Matching via Memory-Efficient Data Packing and In-Flash Processing}
\title{\Large{\thetitle}}

\author{Mayank Kabra$\dagger$ \quad
Rakesh Nadig$\dagger$ \quad 
Harshita Gupta$\dagger$ \quad 
Rahul Bera$\dagger$ \quad 
Manos Frouzakis$\dagger$ \quad 
\\
Vamanan Arulchelvan$\dagger$ \quad 
Yu Liang$\dagger$ \quad 
Haiyu Mao$\ddagger$ \quad 
Mohammad Sadrosadati$\dagger$ \quad 
Onur Mutlu$\dagger$\vspace{-3mm}\\\\
\emph{ETH Zurich}$\dagger$ \qquad \emph{King's College London}$\ddagger$}

\begin{document}
\maketitle

\input{sections/00-abstract}

\input{sections/01-introduction} 
\input{sections/02-background}
\input{sections/03-overview}

\input{sections/04-implementation}
\input{sections/05-methodology}

\input{sections/06-evalutation}
\input{sections/07-discussion}
\input{sections/08-related-work}
\input{sections/09-conclusion}

\bibliographystyle{IEEEtran}
\bibliography{refs}

\end{document}

%% file: packages.tex
\usepackage{amsmath,amssymb,amsfonts}

\usepackage{graphicx}
\usepackage{textcomp}

\usepackage{rotating}
\usepackage{pdflscape}
\usepackage{mathptmx} 
\usepackage{siunitx}
\usepackage{multirow}
\usepackage{glossaries}
\usepackage{adjustbox}
\usepackage{tablefootnote}
\usepackage{subcaption}
\usepackage{setspace}
\usepackage{balance}
\usepackage{datetime}
\usepackage{dblfloatfix}

\usepackage{lipsum}
\usepackage{makecell}

\usepackage[bookmarks=true,breaklinks=true,letterpaper=true,colorlinks,citecolor=blue,linkcolor=blue,urlcolor=blue]{hyperref}

\usepackage[sort,compress]{cite}

\usepackage{xcolor, soul}
\usepackage{xspace}
\usepackage{tabularx}
\usepackage{tikz}
\usepackage[bottom]{footmisc}
\newcommand*\circled[1]{\tikz[baseline=(char.base)]{\node[shape=circle,fill,inner sep=1pt] (char) {\textcolor{white}{#1}};}}
\usepackage[most]{tcolorbox}
\usepackage{pifont}

\renewcommand{\ttfamily}{\rmfamily}

\usepackage{algorithm}
\usepackage{algpseudocode}

\usepackage{fancyhdr}
\usepackage{mathtools}


%% file: commands.tex
\definecolor{darkspringgreen}{rgb}{0.09, 0.45, 0.27}
\definecolor{denim}{rgb}{0.08, 0.38, 0.74}
\definecolor{darkolivegreen}{rgb}{0.33, 0.42, 0.18}
\definecolor{tangerine}{rgb}{0.95, 0.52, 0.0}
\definecolor{mahogany}{rgb}{0.75, 0.25, 0.0}
\definecolor{coolblack}{rgb}{0.0, 0.18, 0.39}

\definecolor{seagreen}{rgb}{0.18, 0.55, 0.34}

\newcommand*\circledd[1]{%
    \tikz[baseline=(char.base)]{%
        \node[shape=circle,fill,inner sep=1pt, minimum size=11pt, font=\bfseries\normalsize] (char) {\textcolor{white}{\footnotesize #1}};%
    }%
}

\newcommand*\circleddcolor[1]{\tikz[baseline=(char.base)]{
    \node[shape=circle, draw=black, fill=white, inner sep=1.0pt, font=\bfseries\normalsize] (char) {\textcolor{black}{#1}};
}}

\newcommand{\harshita}[1]{\textcolor{black}{#1}}
\newcommand{\harshitaaa}[1]{\textcolor{black}{#1}}

\newcommand{\mayank}[1]{\textcolor{black}{#1}}

\newcommand{\mayankk}[1]{\textcolor{black}{#1}}

\newcommand{\mkcr}[1]{\textcolor{black}{#1}}

\newcommand{\mk}[1]{\textcolor{black}{#1}}

\newcommand{\mkk}[1]{\textcolor{black}{#1}}

\newcommand{\mkkk}[1]{\textcolor{black}{#1}}

\newcommand{\mkv}[1]{\textcolor{black}{#1}}

\newcommand{\mkvv}[1]{\textcolor{black}{#1}}

\newcommand{\mkvvv}[1]{\textcolor{black}{#1}}

\newcommand{\mkn}[1]{\textcolor{black}{#1}}

\newcommand{\mknf}[1]{\textcolor{black}{#1}}

\newcommand{\mknn}[1]{\textcolor{black}{#1}}

\newcommand{\mknnn}[1]{\textcolor{black}{#1}}

\newcommand{\mknnnn}[1]{\textcolor{black}{#1}}

\newcommand{\mkf}[1]{\textcolor{black}{#1}}

\newcommand{\orphan}[1]{\textcolor{black}{#1}}

\newcommand{\change}[1]{\textcolor{black}{#1}}

\newcommand{\mkneww}[1]{\textcolor{black}{#1}}

\usepackage{chngcntr}
\usepackage{caption}

\newcommand{\mayankkk}[1]{\textcolor{black}{#1}}

\newcommand{\manos}[1]{\textcolor{black}{#1}}

\definecolor{cerulean}{rgb}{0.0, 0.48, 0.65}

\newcommand{\tklbl}{Key Takeaway}
\newcounter{take}
\newcommand\takeaway[1]{%
    \refstepcounter{take} 
    \noindent 
    \colorbox{cyan!20}{\textbf{\tklbl{} \thetake.}} 
    \textit{#1}\par\medskip 
}

\newcommand{\revision}[1]{\textcolor{black}{#1}}

\definecolor{darkgreen}{rgb}{0.0, 0.7, 0.0} 

\definecolor{darkblue}{rgb}{0.0, 0.0, 0.7} 
\newcommand{\writing}[1]{\textcolor{black}{#1}}

\newcommand{\rn}[1]{\textcolor{black}{#1}}

\newcolumntype{Y}{>{\centering\arraybackslash}X}
\usepackage{tikz}

\newcommand{\squishlist}{
 \begin{list}{$\circ$}
  { \setlength{\itemsep}{0pt}
     \setlength{\parsep}{0pt}
     \setlength{\topsep}{3pt}
     \setlength{\partopsep}{0pt}
     \setlength{\leftmargin}{1em}
     \setlength{\labelwidth}{1em}
     \setlength{\labelsep}{0.5em} } }

\newcommand{\squishend}{
  \end{list}  }

\usepackage{todonotes}
\usepackage{blindtext,graphicx}
\usepackage[absolute]{textpos}
\usepackage{datetime}

%% file: sections/00-abstract.tex
\begin{abstract}
Homomorphic encryption (HE) allows secure computation on encrypted data without revealing the original data, \writing{providing significant} benefits for privacy-sensitive \writing{applications}. Many
cloud computing applications (e.g., DNA read mapping, biometric matching, web search) use \mknf{exact} string matching as a key operation. However, prior \rn{string matching algorithms that use} homomorphic encryption \rn{are limited by} high computational latency \mkv{caused by the use of complex operations} and data movement bottlenecks due to the large encrypted data size. \writing{In this work, we} \mk{provide} an efficient algorithm-hardware codesign \mkv{to} accelerate \mkv{HE-based secure exact string matching}. \writing{\mk{We} propose CIPHERMATCH, which (i) \mkv{reduces} the \mkk{increase in memory footprint after encryption} using an optimized \mk{software-based} \mkcr{data} packing scheme, \mkk{(ii) eliminates the use of costly homomorphic operations (e.g., multiplication and rotation),} and (iii) reduces data movement by designing a new in-flash processing (IFP) architecture.}

\mayank{CIPHERMATCH \rn{improves the \mkv{software-based data} packing \mkv{scheme} of an existing HE scheme \writing{and}}
perform\mkcr{s} secure string matching using \emph{\mkkk{only}} homomorphic addition. \writing{This packing method} reduces the \mkv{memory footprint} after encryption and improves the performance of the algorithm. To reduce the data movement overhead, we design an IFP architecture to accelerate \mkcr{homomorphic} addition by leveraging the \mkcr{array-level and bit-level} parallelism of NAND-flash-based solid-state drives \writing{(SSDs)}.
\rn{We demonstrate the benefits of CIPHERMATCH using} two case studies: (1) Exact DNA \mkcr{string matching} and (2) encrypted database search.} 
\mk{Our \mk{pure software-based} CIPHERMATCH implementation \mk{that uses our \mkvv{memory-efficient} data packing \mkv{scheme} improves performance and reduces energy consumption by} 42.9$\times$ and 17.6$\times$, respectively, compared to the state-of-the-art software baseline. \mk{Integrating CIPHERMATCH with IFP improves performance and reduces energy consumption by 136.9$\times$ and  256.4$\times$, respectively, compared to the software-based CIPHERMATCH implementation.}}
\end{abstract}

%% file: sections/01-introduction.tex

\section{Introduction}
\mkcr{\textbf{Homomorphic Encryption (HE)}~\cite{ogburn2013homomorphic, tourky2016homomorphic, gentry2011implementing, al2020towards, gentry2009fully, van2010fully, cryptoeprint:2014/356, boneh2013private,fan2012somewhat} offers a solution}
\mkcr{to perform computation directly on encrypted data without the need for decryption~\cite{moore2014practical, chaudhary2019Analysis, gentry2009fully}.} \mkv{This capability is essential for processing large volumes of security-sensitive data (e.g., DNA sequences~\cite{organick2020probing,cali2020genasm, bhukya2011exact, firtina2023rawhash, de2013secure, aziz2024secure}, biometrics~\cite{verma2019hybrid,chen2013face}, and web application databases~\cite{mr2016homomorphic,silberschatz1997database, koudas2004flexible, raj2020web}) in shared computing environments. Many applications \mkcr{perform exact string matching on security-critical data, such as DNA string matching~\cite{cali2020genasm, bhukya2011exact, firtina2023rawhash, de2013secure, aziz2024secure}, biometric signature matching~\cite{chen2013face}, and privacy-preserving database searches~\cite{silberschatz1997database, koudas2004flexible, raj2020web}}. However, to perform the string matching operation \mk{in conventional systems (without HE)}, the data must be decrypted before computation (or stored in an unencrypted format). Since 
cloud servers can be shared among multiple users, \mk{sensitive} user data can become vulnerable to security threats and leaks~\cite{shaikh2016attacks, duncan2015overview, bella2022study}.}
HE can significantly benefit privacy-sensitive applications~\cite{yasuda2013secure, essex2019secure, bonte2020homomorphic, yasuda2014privacy, feer2024privacy} that require exact string matching
~\cite{silberschatz1997database, koudas2004flexible, raj2020web, chen2013face, cali2020genasm, bhukya2011exact, firtina2023rawhash, de2013secure, aziz2024secure} 
as the fundamental operation \mk{by directly \mkkk{operating} on encrypted data \orphan{without requiring decryption}.}
\mayankk{\mk{Unfortunately,} homomorphic operations
are typically \mkkk{$10^4\times$} to $10^5\times$ slower than their traditional unencrypted counterparts \mk{in existing systems}~\cite{feldmann2021f1}.}
\mkk{Prior works propose two main approaches to perform secure string matching: (1) the Boolean approach (e.g.,~\cite{aziz2024secure, pradel2021privacy}), and (2) the arithmetic approach (e.g.,~\cite{yasuda2013secure, kim2017private, bonte2020homomorphic}). The Boolean approach~\cite{aziz2024secure, pradel2021privacy} packs individual bits into a polynomial, encrypts it, and uses homomorphic XNOR and AND operations to perform secure string matching on \mkkk{a} search pattern of any size. In contrast, the arithmetic approach~\cite{yasuda2013secure, kim2017private, bonte2020homomorphic} packs multiple bits into a polynomial, encrypts it, and employs homomorphic multiplication and addition operations to compute the Hamming Distance (HD)~\cite{bookstein2002generalized} for approximate or exact secure string matching on search patterns of only specific sizes (see \S\ref{sec:ssm} \orphan{for more detail}).}

\mkk{Despite recent \writing{performance improvements in \mkv{accelerating}}
HE operations (e.g.,~\cite{kim2022bts,feldmann2021f1,samardzic2022craterlake}),} 
\mkk{we make two key observations from \mkf{prior works on secure string matching}~\cite{aziz2024secure,pradel2021privacy,yasuda2013secure,kim2017private,bonte2020homomorphic} that highlight \writing{opportunities for significant  \mk{performance improvement}. \mkk{ First, prior works
\mkv{that employ} the Boolean approach incur substantial latency overheads due to the need to traverse large homomorphically encrypted
databases (e.g.,~\cite{aziz2024secure, petrean2024homomorphic}). In contrast, \mkkk{the arithmetic approach} achieve\mkkk{s} a lower memory footprint after encryption using data packing \mkv{schemes}, but \mkkk{relies} on costly homomorphic operations, such as homomorphic multiplication and rotation (see \S\ref{sec:lim} for \mkkk{more} detail). However, the data packing schemes and algorithms used in \mkkk{the} arithmetic approach can be further optimized \mkf{by using} simpler operations (e.g., homomorphic addition) to \mkkk{significantly} improve the performance of secure \mkkk{exact} string matching.}}}

\mkk{{{Second, a significant portion of the performance overhead in HE comes from the movement of large homomorphically encrypted databases, which can be up to 1000$\times$ larger than their unencrypted counterparts~\cite{feldmann2021f1, samardzic2022craterlake, gupta2023evaluating}\mk{,} across the memory hierarchy. \harshita{Several near-data processing approaches\mk{~\cite{mutlu2022modern,nider2020processing}, such as (1) processing-near memory (PnM) (e.g.,~\cite{farmahini2014drama, singh2018review, kwon2019tensordimm, gupta2024memfhe, yitbarek2016exploring,ahn2015scalable,ahn2015pim}), (2) processing-using memory 
(PuM) (e.g.,~\cite{hajinazar-asplos-2021, seshadri2017ambit, ferreira2021pluto, wang2020figaro}), (3) in-storage processing (ISP) (e.g.,~\cite{gu2016biscuit, jo-vldb-2016, ghiasi2022genstore, wang2016ssd, park2016storage, ruan2019insider, wang2016ssd1}) and (4) in-flash processing (IFP) (e.g.,~\cite{park2022flash, chen2024ares, gao2021parabit})} have been proposed to address the data movement bottleneck \mayank{between storage \mk{and} main memory \mk{system} and \mk{the compute-centric processors or accelerators}}}. While these solutions \harshita{aim to} utilize high parallelism and \mkkk{high} memory bandwidth of the storage \mk{and} main memory devices, 
\writing{prior secure string matching techniques~\cite{aziz2024secure,pradel2021privacy,yasuda2013secure, kim2017private,bonte2020homomorphic} have limitations \mkf{in} the data packing schemes\mk{,} \mkkk{resulting in large memory footprint leading} to inefficient use of the existing \mkkk{near-data processing} architectures\mk{, \mkv{and} compute-centric systems} (see \S\ref{sec:lim}).} }}}

\textbf{Our goal} in this work is to improve \mkkk{the} performance of \mkv{homomorphic encryption-based secure} string
matching \harshita{by \mkcr{providing} an efficient algorithm-hardware codesign.}
\harshitaaa{To this end, we propose} 
\mayankk{CIPHERMATCH\footnote{CIPHERMATCH is derived from a combination of \underline{cipher}text from homomorphic encryption and string \underline{match}ing, reflecting its core functionality in secure string matching on encrypted data.}, whose \textbf{key idea} is to reduce \mkf{(1) the} \mkk{increase in memory footprint after encryption}  \mkk{by optimizing the \mkv{existing data packing schemes} used by \mkkk{the} arithmetic approach, (2) computational latency (\mkkk{i.e.}, execution time) of secure exact string matching by eliminating costly homomorphic multiplication or rotation operations, and (3) data movement overhead by moving computation \mkkk{into the} storage \mkkk{devices} using a new \mkkk{in-flash processing design}}.}
\mayankk{\textbf{Key Mechanisms.}}
To enable CIPHERMATCH, we employ four key mechanisms. First, we develop \textbf{\mk{a memory-efficient} \mkcr{data} packing scheme} that packs \emph{multiple bits} of the database and \mk{search pattern into a polynomial} before encryption, \mk{reducing} the \mk{memory footprint} of the encrypted database and encrypted search pattern. Second, we \writing{implement secure \string matching using} \mkkk{\emph{only}} \mk{the} homomorphic addition operation (\mk{an optimization enabled by our} efficient \mkcr{data} packing \mkv{scheme}), which reduces the execution time \mk{of secure string matching \mkv{algorithm}}. Third, we \mkkk{use in-flash processing and} implement homomorphic addition directly within NAND-flash chips using read and latch operations~\cite{park2022flash, gao2021parabit}, \mkkk{utilizing} the inherent \mkcr{array-level and bit-level} parallelism \mk{in flash chips}. \mkk{Fourth, we utilize this in-flash processing architecture to enable CIPHERMATCH in \mkkk{\mkf{a} modern SSD-based storage system} (called \emph{CM-IFP}).}

\mkk{\textbf{Key Results.} 
\mkk{We demonstrate the benefits of CIPHERMATCH
using two case studies: (1) \mkvvv{exact} DNA string matching and (2) encrypted
database search. We compare \mkkk{our} pure software-based CIPHERMATCH implementation (CM-SW) that uses our \mkvv{memory-efficient} data packing scheme on a real CPU system \mkk{(see \S\ref{sec:met})} with \mkkk{a state-of-the-art} \mkkk{software baseline}} ~\cite{yasuda2013secure} \mkk{(see \S\ref{sec:ssm}).} 
Our evaluation shows that \mkkk{CM-SW} improves performance and reduces energy consumption by 42.9$\times$ and 17.6$\times$, respectively, \mkkk{over the state-of-the-art.}
\mkk{To assess the performance of CIPHERMATCH with IFP (CM-IFP), we compare CM-IFP with (1) CM-SW and (2) CIPHERMATCH implementation using processing-using memory (CM-PuM)~\cite{hajinazar-asplos-2021}. \mkkk{Our results demonstrate that CM-IFP improves performance over CM-SW and CM-PuM by 136.9$\times$ and 1.4$\times$, and reduces energy consumption by 256.4$\times$ and 3.3$\times$, respectively.}
}}

\noindent This work makes the following \textbf{key contributions}:
\squishlist
    \item \harshita{We propose CIPHERMATCH, a \mkv{new} approach that improves the performance \mk{and energy efficiency} of \mkkk{homomorphic encryption \orphan{(HE)} based} secure \mknf{exact} string matching.}
    \item \harshita{CIPHERMATCH \mk{introduces} a new \mkvv{memory-efficient} \mkcr{data} \mayankk{packing} scheme that \mkvvv{drastically} reduces the \mkvvv{large} \mk{increase in memory footprint} after encryption, leading to faster \mk{computation} and reduced storage overhead \mk{for HE operations. \mkkk{The proposed data packing scheme improves the \mkvvv{performance and} efficiency of \mkvvv{both} compute-centric and near-data processing architectures by increasing parallelism and reducing data movement overhead of secure string matching.}}}
    \item 
    \mk{CIPHERMATCH} \harshita{is the first work to 
    \mk{exploit} in-flash processing for HE operations, enabling efficient secure string matching directly on encrypted data within the SSD. This approach reduces the data movement bottleneck and leverages the \mkcr{array-level and bit-level parallelism} of flash memory.}
    \item \mk{Both pure software-based and \mkvvv{in-flash processing} implementations of CIPHERMATCH significantly improve performance \mkkk{and energy efficiency} \mkkk{over} state-of-the-art \mkkk{software and hardware} approaches on real-world applications like DNA string matching and encrypted database search.}
\squishend





%% file: sections/02-background.tex
\section{Background}

\subsection{Homomorphic Encryption}
\label{sec:he}
Homomorphic encryption (HE)~\cite{ogburn2013homomorphic, tourky2016homomorphic, gentry2011implementing, al2020towards, gentry2009fully, van2010fully, cryptoeprint:2014/356, boneh2013private,fan2012somewhat} is an encryption \mk{paradigm} that allows users to compute directly \mayank{on} encrypted data (ciphertext) data without decrypting it. Most HE schemes build upon the 
\mk{\emph{learning with errors (LWE)}} problem~\cite{Regev:2005aln,lyubashevsky2010ideal}.
In the context of HE schemes, encrypted plaintexts can be typically represented in vector format, leading to highly parallel computation
~\cite{smart2014fully,castryck2018homomorphic,qin2019addition}. In this work, we utilize the Brakerski-Fan-Vercauteren (BFV)~\cite{fan2012somewhat} HE scheme.

\textbf{Brakerski-Fan-Vercauteren (BFV) HE Scheme.}
The \mk{BFV} scheme~\cite{fan2012somewhat}  is built on the Ring-LWE problem~\cite{lyubashevsky2010ideal}, which leverages polynomial rings to compute on encrypted data. The BFV scheme operates within a ring structure 
R = {$\mathbb{Z}_q$}[X]/{($X^n$+1)}, 
where (1) $n$ (ring dimension) denotes the maximum polynomial degree, (2) q is the coefficient bit-length of ciphertext, and (3) $\mathbb{Z}_q$ denotes a set of integers modulo $q$. Arithmetic operations within this ring are performed modulo $X^n$ + 1. The plaintext space  $R_t$ consists of polynomials ( $t$ $\geq$ $2$ ), where t is the coefficient bit length of plaintext and  ($X^n$+1), while the ciphertext space $R_q$ utilizes a larger coefficient modulus ( $q$ $\gg$ $t$ ). \mk{These parameters (n, q, t) are defined \mk{according to the} security level of \mk{the} homomorphic encryption scheme~\cite{HomomorphicEncryptionSecurityStandard}}. The BFV scheme leverages \mk{several algorithms (e.g., key generation, encoding, encryption, and homomorphic operations)~\cite{gentry2009fully}} to achieve its functionalities. 

\revision{\textbf{Key Generation.} A user generates a secret key $sk \in \mathbb{P}_{n} \left(\mathbb{Z}_{q}\right)$ and a public key pair $\left(pk_0, pk_1\right) \in \mathbb{P}^{1}_{n} \left(\mathbb{Z}_{q}\right)$ based on the required security parameters. $\mathbb{P}_n\left(R\right)$ denotes the set of polynomials of degree less than $n$ with coefficients in ring $R$.}

\revision{\textbf{Encoding and Encryption.} A message $m \in \mathbb{Z}_{t}$ is converted to a plaintext polynomial $M(x) \in \mathbb{P}_{n} \left(\mathbb{Z}_{t}\right)$ using an encoding scheme. The encryption function takes the public key \mk{pair} $\left(pk_0, pk_1\right)$ and a plaintext polynomial $M\left(x\right)$ and outputs a ciphertext C = ($C_0$, $C_1$).}
\begin{equation}
\orphan{C = \left(C_0\left(x\right), C_1\left(x\right)\right) \in \left(\mathbb{Z}_q[x], \mathbb{Z}_q[x]\right)}
\end{equation}
\begin{equation}
\orphan{C_0\left(x\right) = [pk_0\left(x\right) + e_0\left(x\right) +M\left(x\right)]_q}
\end{equation}
\begin{equation}
\orphan{C_1\left(x\right) = [pk_1\left(x\right) + e_1\left(x\right)]_q}
\end{equation}
\orphan{$e_0\left(x\right)$ and $e_1\left(x\right)$ are polynomials introduced to achieve the 
security properties of the scheme}.

\revision{\textbf{Homomorphic Addition \textit{(Hom-Add)}.}
The addition between two ciphertext $C^1 = (C_0^1, C_1^1)$ and $C^2 = (C_0^2, C_1^2)$ is performed coefficient-wise on the corresponding polynomials in the ciphertexts generated after encryption, \mk{as shown in Eq.~\eqref{eq:encryption}:}}
\revision{\begin{equation}
\label{eq:encryption}
\orphan{C\left(x\right) = \left( C_0^1\left(x\right) \text{+ } C_0^2\left(x\right), \, C_1^1\left(x\right) \text{+ } C_1^2\left(x\right) \right)}
\end{equation}}


\subsection{Secure String Matching Algorithm}
\label{sec:ssm}
\mayankk{\mk{The conventional} string matching operation is \mk{widely} used in \mk{various} applications and is \mk{typically} implemented with bitwise XNOR and AND operations\mk{~\cite{olson2020approach, sedghi2010searching}}. Database} systems~\cite{silberschatz1997database,raj2020web,kim2021efficient, koudas2004flexible} \mk{commonly} use \mknf{exact} string matching to search \mk{a} key (i.e., \mk{a} given string) in \mkkk{a} database (i.e., \mk{many} entries consisting of key-value pairs). \mk{Various bioinformatics} applications ~\cite{cali2020genasm,firtina2023rawhash,bhukya2011exact,de2013secure, alser2020accelerating, alser2022molecules,aziz2024secure} (e.g., \mkkk{read mapping} and alignment) map a genome to a reference genome to \mk{identify} genetic variations\mk{, using a combination of exact and approximate string matching}. \mk{These applications can be used in a privacy-sensitive context and thus can} have strict data security rules for user data. To ensure data \mkkk{privacy and} security, these applications\mk{, likely} need to perform computations on encrypted data. 

\orphan{HE offers significant benefits for secure string search:
1) Low communication complexity:  HE requires only two rounds of data exchange between client and server during computation and incurs minimal data transfer overhead compared to other approaches, such as Yao’s garbled circuit~\cite{yao, snyder2014yao}. 2) Non-interactiveness: Unlike multi-party computation (MPC)~\cite{snyder2014yao, goldreich2019play}, HE does not require continuous interaction of users during computation, reducing the need for extensive online communication.
3) Universality: HE allows the implementation of any string matching algorithm without extensive data preprocessing, making the data accessible for other computational tasks, unlike other approaches (e.g., ~\cite{chor1998private,goldreich1987towards,chen2018labeled}).
4) No data leakage: HE is designed to hide 
all information about encrypted data, except for maximum data size, making it a secure option for string matching on encrypted data. While symmetric searchable encryption-based string matching schemes (e.g.,~\cite{chase2015substring,leontiadis2018storage}) can be very efficient in terms of
storage overhead and computation speed, their leakage
profiles tend to be quite substantial~\cite{poh2017searchable, giraud2017practical,grubbs2017leakage}. This has led to various
attacks that manage to recover parts of the plaintext or the queries~\cite{grubbs2017leakage}. Hence, in this work, our focus is to leverage HE for secure string matching.} There are two key approaches to perform\mkkk{ing} string matching using HE.


\textbf{Boolean Approach} encrypts the individual bits of the database and the query using the TFHE (Fast Fully Homomorphic Encryption over the Torus) encryption scheme~\cite{chillotti2020tfhe,TFHE-rs}. 
In this context, homomorphic XNOR and homomorphic AND operations are performed on the encrypted database with the encrypted query. This method requires traversing an encrypted database and performing homomorphic operations on individual encrypted bits of the database and query. Each bit is encrypted and converted to an encrypted polynomial. 
The Boolean approach effectively limits noise growth in homomorphic encryption, enabling an arbitrary number of computations~\cite{chillotti2020tfhe}. \change{As a result, it supports string matching for queries of any length, providing users with flexibility in designing string matching algorithms.}

\mkk{\textbf{Arithmetic Approach} encrypts the database and the query using SHE (Somewhat Homomorphic Encryption)-based schemes~\cite{yasuda2013secure,yasuda2015new,kim2017private,bonte2020homomorphic}. In this context, we encode both an input database and a query as integers and follow a three-step procedure. First, we partition the integer $M$, where $M$ is at most `k' bits, \mkf{and convert into} a binary vector ($M_{0}$,...,$M_{k-1}$) 
which is then expressed as a polynomial P(M) = $\Sigma_{i=0}^{k-1}M_ix^i$.
Second, we represent P(M) as our new message and encrypt the polynomial (see \S \ref{sec:he}). Third, we calculate the Hamming Distance (HD)~\cite{bookstein2002generalized} between encrypted database and encrypted query using homomorphic multiplication and addition. This method avoids encrypting individual bits separately and instead packs multiple bits (\mkkk{i.e., performs} data packing) within a single ciphertext, significantly reducing the overall data size. However, this data packing scheme restricts the query size encoded within the plaintext polynomial, as SHE permits only a finite number of computations on encrypted data~\cite{fan2012somewhat}. Hence, string matching can only be performed on specific query lengths. \change{We discuss the limitations of the state-of-the-art techniques using this data packing method in \S\ref{sec:lim}.}}

\subsection{Basics of NAND Flash Architecture}
\label{sec:NANDflash}

\begin{figure}[h]
    \centering
    \includegraphics[width=\linewidth]{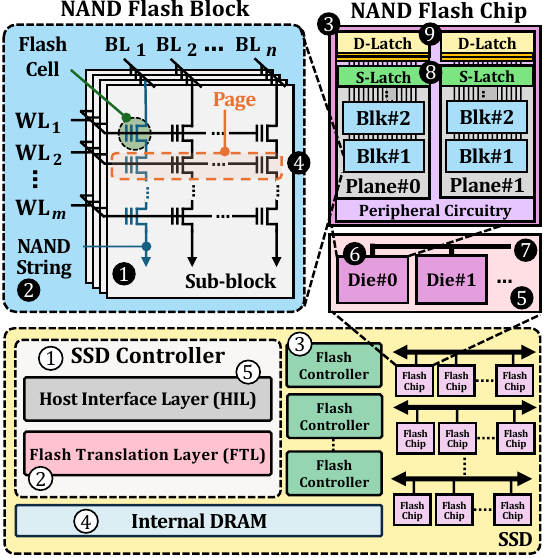}
    \caption{High-level overview of \mkkk{a} modern SSD.}
    \label{fig:nand}
\end{figure}

\mkk{Figure \ref{fig:nand} shows the internal organization of a modern 3D NAND-flash based SSD.}

\mkk{\textbf{NAND Flash Organization.} Multiple floating-gate transistors are stacked serially, forming a NAND string~\cite{micheloni-insidenand-2010,micheloni2018inside}. A NAND string is connected to a bitline (BL), and NAND strings \mk{connected to} multiple BLs compose a sub-block \circled{1}. The control gates of all cells at the same vertical position within a sub-block are linked to a single wordline (WL), allowing for the concurrent operation of these cells. A NAND flash block \circled{2} consists of several (e.g., 4 or 8) sub-blocks, and thousands of blocks comprise a plane \circled{3}. A block includes hundreds to thousands of pages \circled{4}, each of which is 4–16 KiB in size. \change{The blocks in a plane share all the BLs in that plane, which implies that thousands of NAND strings share a single BL.} \mkk{A NAND flash chip \circled{5} contains multiple (e.g., 2 or 4) dies \circled{6}, and each die contains multiple (e.g., 2 or 4) planes. Multiple dies can operate independently but share the command/data buses (i.e., channel \circled{7}) in a time-interleaved manner.}}

\mkk{\textbf{NAND Flash Operations and Peripheral Circuitry.} 
\mkkk{Modern NAND flash memory operates in single-level (SLC)~\cite{cho2001dual}, multi-level (MLC)~\cite{lee20043}, triple-level (TLC)~\cite{maejima2018512gb}, or quad-level (QLC)~\cite{cho20221} modes, where each cell stores 1, 2, 3, or 4 bits, respectively.}}
\mkkk{In modern NAND flash memory, during a read operation~\cite{cai2012flash,cai2015read,cai2015data,cai2013program,cai2013threshold,li2018access} an initial precharge voltage ($V_{pre}$) charges the bitline, and a read voltage ($V_{read}$) is supplied to the targeted wordline. If the cell has a high threshold voltage ($V_{th}$), the bitline remains in the logical $1$ state; otherwise, \change{the bitline} \mkvvv{is pulled down} to logical $0$. During this phase, the bitline value is latched into the sensing latch \circled{8} and sent to the SSD controller via flash channels. Modern NAND flash memory has a sensing latch and multiple (e.g., 3 or 4) data latches ~\circled{9} per plane \cite{micheloni-insidenand-2010}, \mkvvv{to buffer data read from multiple bits stored in a single cell}\mk{~\cite{cai2017vulnerabilities}}. \change{We explain in detail the functionality of both data latches (D-latches) and sensing latch (S-latch) in \S\ref{sec:ifp}.}}

\mkkk{\textbf{Modern SSD Organization}.
Modern SSDs consist of four key components: 1) the SSD controller~\circleddcolor{1}, which includes multiple embedded cores to manage the flash translation layer (FTL)~\circleddcolor{2}; 2) per-channel hardware flash memory controllers~\circleddcolor{3} for request handling~\cite{tavakkol2018flin, cai2013program}; 3) DRAM \orphan{(e.g., 2GB LPDDR4 DRAM for a 2TB SSD)}~\circleddcolor{4}~\cite{samsung-980pro} for caching logical-to-physical (L2P) mappings and storing metadata; and 4) the host interface layer (HIL)~\circleddcolor{5} for communication between the host system and SSD. The FTL is responsible for I/O scheduling, address translation, and garbage collection and uses DRAM to cache a portion of L2P mappings at sub-byte granularity (with the memory overhead of \textasciitilde0.1\% of the SSD capacity), while the HIL handles I/O requests through AHCI~\cite{sata} or NVMe~\cite{NVMe2021} protocols. }

\subsection{Near-Data Processing}
\label{sec:pim}

Near-data processing (NDP)~\cite{mutlu2022modern,nider2020processing} offers a promising solution to meet the high bandwidth requirements of memory-intensive workloads by performing computation where the data resides\mk{~\cite{mutlu2019processing}}. NDP can be categorized into \mk{two approaches:} (1) processing-in (main) memory (PIM), \mkkk{which} enables the memory subsystem to perform computation and (2) processing-in storage, \mkkk{which} enables the storage devices to perform computation. PIM can be further categorized into two approaches: processing-using memory (PuM) and processing-near memory (PnM)~\cite{mutlu2022modern}. Processing-in storage can be further categorized into two approaches: in-storage processing (ISP) and in-flash processing (IFP)~\cite{nider2020processing}.

\mkk{Processing-using memory (PuM) approaches \mk{(e.g.,~\cite{hajinazar-asplos-2021, seshadri2017ambit, ferreira2021pluto, wang2020figaro})} use the \mkkk{operational principles of memory circuitry and chips} to perform operations within the main memory, while processing-near memory (PnM) approaches \mk{(e.g.,~\cite{gupta2024memfhe, farmahini2014drama, singh2018review, kwon2019tensordimm, yitbarek2016exploring, ahn2015scalable, ahn2015pim})} \mkkk{add} processing elements inside or near the \mk{main} memory to perform computations with reduced data movement overhead. However, with the increasing \mkk{dataset sizes (e.g.,~\cite{ghiasi2024megis, ghiasi2022genstore})}, \mkk{PIM approaches are increasingly bottlenecked by storage (SSD) accesses (as the \mkkk{entire} dataset is typically stored in storage devices)~\cite{yang2023end}. With limited I/O bandwidth and high latency of SSD accesses, a \mkk{significant amount of} time is spent loading the \mkk{application} data from storage to main memory~\cite{ghiasi2022genstore,ghiasi2024megis,park2022flash,gao2021parabit,chen2024ares}.}}

In-storage processing (ISP) approaches \mk{(e.g.,~\cite{gu2016biscuit, jo-vldb-2016, wang2016ssd, park2016storage, ruan2019insider, wang2016ssd1})} offload tasks to an in-storage accelerator \mk{or embedded cores \mkv{in an SSD controller}}, which can significantly accelerate and improve the performance of data-intensive applications \mk{(e.g.,~\cite{wong2024tcam, fakhry2023review,park2014query,kang2013enabling,koo2017summarizer})} by leveraging high internal \mk{SSD} bandwidth to perform computation. 

In-flash processing (IFP) approaches \mk{(e.g.,~\cite{gao2021parabit,park2022flash,shim2022gp3d,chen2024ares})} offer performance \mk{and energy} benefits \mk{over ISP} by \mkv{exploiting} \mk{the operational principles of NAND} flash cells \mk{and circuitry} for \mkv{performing massively parallel }computation, focusing on applications that use bulk-bitwise operations \mk{(e.g.,~\cite{chan-signmod-1998, oneil-ideas-2007, li-vldb-2014, li-sigmod-2013, goodwin-SIGIR-2017, seshadri-micro-2013, seshadri2017ambit, seshadri-ieeecal-2015, hajinazar-asplos-2021, FastBitA9, wu-icde-1998, guz-ndp-2014,redis-bitmaps,alser-bioinformatics-2017, loving-bioinformatics-2014, xin-bioinformatics-2015, cali2020genasm, kim-genomics-2018, lander-nature-2001, altschul-jmb-1990, myers-jacm-1999})}. 

\change{Several NDP-based approaches have been proposed to accelerate homomorphic encryption (HE) by leveraging the massive parallelism and low-latency memory access of NDP architectures~\cite{gupta2024memfhe,reis2020computing,gupta2023evaluating,zhou2025fhemem,mwaisela2024evaluating,yang2023poseidon,suzuki2023designing}. We discuss the limitations of these prior works in \S\ref{sec:accel}.}


\section{Motivation}
\orphan{We provide our motivational analysis in three areas. First, we qualitatively discuss the limitations of prior approaches~\cite{aziz2024secure,pradel2021privacy,yasuda2013secure, kim2017private,bonte2020homomorphic} and quantitatively analyze two prior approaches (Boolean~\cite{aziz2024secure} and arithmetic ~\cite{yasuda2013secure}) that implement secure string matching using HE by comparing their memory footprint and execution time (see \S \ref{sec:lim}). Second, we provide an analysis to demonstrate the data movement overhead in the memory hierarchy for various database sizes (see \S\ref{sec:ndp}). Third, we discuss the limitations of prior HE accelerators and near-data processing architectures for HE (see \S\ref{sec:accel}).}


\subsection{Limitations of Prior \mkkk{Approaches}}
\label{sec:lim}
\mkk{Prior works typically leverage two key approaches to perform secure string matching} \writing{(see \S \ref{sec:ssm}):} (i)~\mkk{Boolean} and (ii) \mkk{arithmetic}. \mkk{To compare these two approaches, we \mkv{examine} their four different characteristics: 1) \mkkk{\emph{execution time}}, the time required for the algorithm to perform the string matching operation; 2) \mkkk{\emph{algorithm scalability}}, the ability of the technique to scale with increasing database sizes; 3) \mkkk{\emph{SIMD support}}, whether the technique can leverage SIMD instructions to enhance performance; and 4) \mkkk{\emph{flexible query size}}, the ability of the technique to support different query sizes}. \mkkk{Table ~\ref{tab:compare} provides a comparison of four different characteristics of \mkv{two different Boolean approaches}~\cite{aziz2024secure,pradel2021privacy} with \mkv{those of three} different arithmetic approaches~\cite{yasuda2013secure,kim2017private,bonte2020homomorphic}.}

\mayankkk{\textbf{Boolean Approach.} \mk{Prior} works \mkv{(e.g., ~\cite{aziz2024secure,pradel2021privacy})}
leverage \mkv{the} TFHE-rs library~\cite{TFHE-rs} to perform \mayankk{homomorphic \mk{Boolean} operations (XNOR and AND) for \mk{secure} string matching.} \mkv{Table ~\ref{tab:compare} provides different characteristics of two state-of-the-art techniques: (1) Pradel et al.~\cite{pradel2021privacy} use the TFHE library to develop a biometric string matching algorithm using basic Boolean homomorphic gates. However, their design does not leverage SIMD batching. (2) Aziz et al.~\cite{aziz2024secure} improve upon~\cite{pradel2021privacy} by implementing SIMD batching to enhance the performance of Boolean homomorphic operations.} However, two major challenges persist across Boolean approaches. \mkk{First, performing a single secure \mknf{exact} string matching operation requires traversing the entire encrypted database and executing computationally expensive homomorphic operations, \mkf{which significantly increase} the execution time}. \mkk{For instance, searching a 32-bit query in a small 32-byte database using HE on a real CPU system (see \S\ref{sec:met}) takes 6.6s, whereas the same search on unencrypted data completes in \mkv{only} 5.9$\mu$s.}  
\mkk{Second, the memory \mk{footprint} after encrypting individual bits (see \S\ref{sec:ssm}) is \mkk{larger} (\mkv{by} more than 200$\times$) \mkv{compared to} \mkv{the memory footprint of} unencrypted data.} }


\textbf{Arithmetic Approach.} Prior works \mk{(e.g.,~\cite{yasuda2013secure, kim2017private,bonte2020homomorphic})} propose different \mk{data} packing \mkv{schemes}
to perform secure string matching \mkv{using the arithmetic homomorphic operations (see \S\ref{sec:ssm})}. 
Table \ref{tab:compare} \mkv{provides}
different characteristics of \mk{three} state-of-the-art
techniques\mkv{:} (1) Yasuda et al.~\cite{yasuda2013secure} utilize a \mk{data} packing method that packs multiple bits in one ciphertext to achieve low latency \mk{for secure string matching operation}. However, it suffers from two potential limitations. First, Hamming Distance (HD)\mk{~\cite{bookstein2002generalized}} is used to find the string matches, which requires two costly homomorphic multiplications and three homomorphic additions \mk{for each comparison}. Second, this approach divides the encrypted database into multiple ciphertexts and returns the same number of ciphertexts \mk{as a result of secure string matching}. \mk{Hence,  this approach is not scalable for large database sizes.} (2) Kim et al.~\cite{kim2017private} address the scalability issue by employing a homomorphic equality (HomEQ) circuit (using \mk{homomorphic} multiplication, rotation, and Frobenius map~\cite{gentry2012better} \mk{operations}) to return only the required results\mk{; however,} it incurs high latency due to expensive \mk{HE} operations. (3) Bonte et al.~\cite{bonte2020homomorphic} improve upon~\cite{kim2017private} by implementing SIMD batching and compression techniques to enhance performance\mk{; however, it still requires performing the computationally expensive homomorphic operations and only support\mkv{s} specific query sizes.} 

\begin{table}[h]
    \centering
    \footnotesize
    \renewcommand{\arraystretch}{1.0} 
    \begin{tabularx}{\linewidth}{|>{\centering\arraybackslash\hsize=1.2\hsize}X|>{\centering\arraybackslash\hsize=0.6\hsize}X||>{\centering\arraybackslash\hsize=1.1\hsize}X|>{\centering\arraybackslash\hsize=1.3\hsize}X|>{\centering\arraybackslash\hsize=1.0\hsize}X|>{\centering\arraybackslash\hsize=0.8\hsize}X|}
        \hline
        \textbf{Approach Type} & \textbf{Prior Work} & \textbf{Execution Time} & \textbf{Algorithm Scalability} & \textbf{SIMD Support} & \textbf{Flexible Query Size} \\
        \hline\hline
        \multirow{2}{*}{\shortstack{Boolean \\ Approach}} 
        & \cite{pradel2021privacy} & High & \checkmark & \ding{55} & \checkmark \\
        \cline{2-6}
        & \cite{aziz2024secure} & High & \checkmark & \checkmark & \checkmark \\
        \hline
        \multirow{3}{*}{\shortstack{Arithmetic \\ Approach}} 
        & \cite{yasuda2013secure} & Low & \ding{55} & \ding{55} & \ding{55} \\
        \cline{2-6}
        & \cite{kim2017private} & High & \checkmark & \ding{55} & \ding{55} \\
        \cline{2-6}
        & \cite{bonte2020homomorphic} & High & \checkmark & \checkmark & \ding{55} \\
        \hline
    \end{tabularx}
    \caption{\mkk{Comparison of prior Boolean \mkkk{and arithmetic} approaches based on \mkkk{(i)} execution time, \mkkk{(ii)} algorithm scalability (i.e., the ability of the technique to scale with increasing database size), \mkkk{(iii)} SIMD support, and (iv) flexible query \mkneww{size} (i.e., the ability of the technique to support different query sizes).}}
    \label{tab:compare}
\end{table}

\begin{figure}[h]
    \centering
    \includegraphics[width=\linewidth]{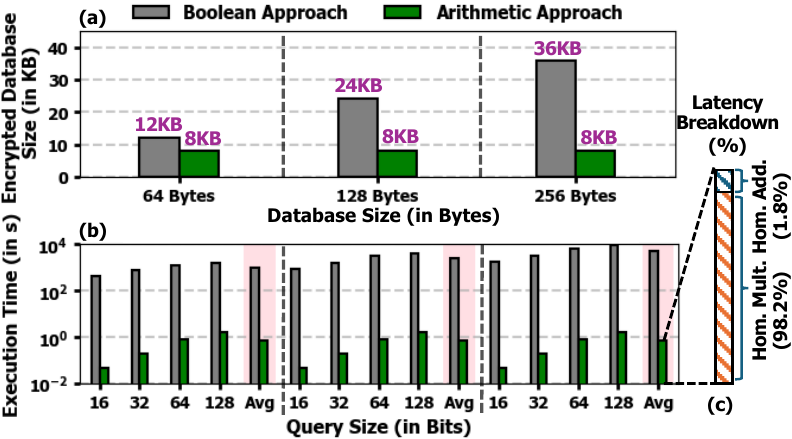}
    \caption{\mkkk{Comparison of \mkvvv{one} Boolean~\cite{aziz2024secure} and \mkvvv{one} arithmetic~\cite{yasuda2013secure} \mkvvv{technique} in terms of (a) memory footprint, (b) execution time (in s), and (c) latency breakdown of the arithmetic approach~\cite{yasuda2013secure}.}}
    \label{fig:motivation}
\end{figure}

\mkk{We \mkkk{quantitatively} evaluate the memory footprint and execution time of two prior works, \mkvvv{one} \mkkk{following the} Boolean approach~\cite{aziz2024secure} and \mkvvv{another,} \mkkk{the} arithmetic approach~\cite{yasuda2013secure}. \mkk{Figure~\hyperref[fig:motivation]{2(a)} shows the memory footprint of the encrypted database (Y-axis) for a given database size (X-axis) for these works. Figure~\hyperref[fig:motivation]{2(b)} shows the execution time (in seconds) for secure string matching across different query sizes (X-axis) for the encrypted database (using the sizes from Figure~\hyperref[fig:motivation]{2(a)}). Figure~\hyperref[fig:motivation]{2(c)} shows the latency breakdown, highlighting the individual contributions of homomorphic addition and multiplication operations.} We use a \mkkk{very} small encrypted database size to understand the \mkv{execution time} of prior works without \mkkk{causing} data movement in the memory hierarchy. We implement \mkkk{the} Boolean approach~\cite{aziz2024secure} using \mkf{the} TFHE-rs library~\cite{TFHE-rs} and \mkkk{the} arithmetic approach~\cite{yasuda2013secure} using the Microsoft SEAL library~\cite{sealcrypto} on a real CPU system (see \S\ref{sec:met}). We make three key observations.}

\vspace{1em}
\mkk{First, the increase in memory footprint introduced \mkkk{due to} encryption is higher for the Boolean approach than the arithmetic approach. This is because the arithmetic approach efficiently groups multiple bits into a single polynomial using \mkkk{a} data packing method before encryption (see \S\ref{sec:he}). In contrast, the Boolean approach encrypts, each bit individually into a polynomial. Second, the execution time (in seconds) of the Boolean approach exceeds that of the arithmetic approach by 600$\times$ when performing secure string matching\mkkk{, averaged} across various \mkkk{database sizes and} query lengths. This is \mkkk{due} to the higher overall cost of executing expensive Boolean homomorphic operations (e.g., XNOR and AND) on \emph{individual encrypted bits}, compared to cheaper arithmetic \mkkk{homomorphic} operations on \emph{multiple packed bits}. Third, 98.2\% of the latency in secure string matching \mkkk{using the arithmetic approach} is due to the expensive homomorphic multiplication operations required to perform secure string matching.}

\noindent\mkk{\takeaway{\label{sec:take1}Data packing scheme\mkv{s} used in arithmetic approaches offer significant performance benefits; \change{however, they can be further optimized \mkkk{to} \mkkk{minimize}} the use of \mkkk{costly} homomorphic multiplication operations and \mkkk{to exploit} highly parallel SIMD units \mkkk{to execute} simple\mkkk{r} homomorphic addition operations.}}

\subsection{Benefits of Near-Data Processing}
\label{sec:ndp}

\mayankk{\mk{Prior works (e.g., \cite{ghiasi2022genstore, ghiasi2024megis}) demonstrate that the traditional string matching applications are constrained by data movement, as the time required for transferring large data volumes across the memory hierarchy significantly impacts performance. Once the databases are encrypted \mkkk{using HE}, their memory footprint \mkkk{greatly} increases (see \S\ref{sec:lim}), exacerbating the bottleneck caused by data movement. To understand the overheads of data movement from storage (SSD) to computation units, we profile}} the data movement overhead associated with transferring data to processing units for different database sizes, ranging from 8GB to 256GB. \mayankk{We assume three scenarios, \mkvvv{where} \mkkk{homomorphic secure string matching computations are performed in:} (1) the CPU, (2) the main memory, and (3) the storage (SSD) controller.}

\begin{figure}[h]
    \centering    \includegraphics[width=0.9\linewidth]{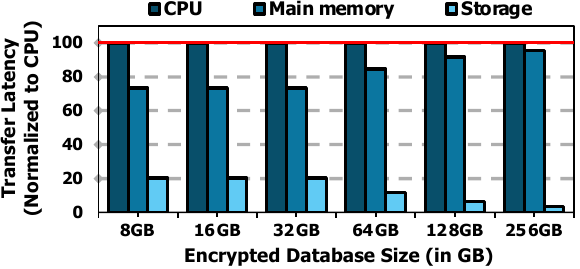}
    \caption{\writing{Transfer latency \orphan{(normalized to the latency of transferring data to the CPU, denoted as Y=100)} for loading data from \mkkk{flash chips} to perform secure string matching \mk{in} (1) CPU; (2) main memory; and (3) storage (SSD) controller.
     \orphan{X-axis denotes the encrypted database size that is transferred.} }}
    \label{fig:data_movement}
\end{figure}

Figure \ref{fig:data_movement} \writing{shows the \mk{data} transfer latency (normalized to \mk{the latency of} \mk{transferring data to the} CPU) from flash memory chips to (i) CPU, (ii) DRAM, and (iii) SSD \mk{controller}.} \mkk{We observe that when the encrypted database is small (e.g., 8GB), \mkkk{performing computations in main memory} lowers data transfer latency by 25\%, \mkkk{because doing so eliminates data movement to the CPU}. As the \mkkk{database} size increases, the overhead of transferring data from \mkkk{flash chips to DRAM becomes more significant, and thus diminishes the latency reduction benefits of performing computations in DRAM}. For all database sizes, performing computations within the SSD controller reduces data transfer latency by over 80\% compared to processing on the CPU. For the largest encrypted database size of 256GB, \mkkk{performing all} computations in the SSD controller results in a 94\% reduction in data transfer latency, whereas performing computations in DRAM provides only a 6\% \mkvvv{latency} reduction.}

\noindent\mkk{\takeaway{\label{sec:take2}As encryption \mk{greatly increases} the data size, applications become increasingly bound by data movement \mkkk{from storage}. Performing computation closer to where the data resides, i.e., \mk{\mkvvv{inside} the SSD controller and flash memory} \mkkk{can greatly} reduce the data movement \mkkk{overheads}.}}

\writing{\textbf{Benefits of In-Flash Processing (IFP).}} 
IFP \mk{enhances} \mk{performance and energy efficiency} by utilizing \mk{the operational principles of} flash \mk{arrays and cells} \mkkk{for massively bit-parallel} computation. Prior approaches \mk{(e.g., \cite{gao2021parabit, park2022flash})} \mk{use NAND flash \mkkk{memory} logic to perform} bulk-bitwise operations since these type of operations are widely used in many applications \mk{(e.g., ~\cite{chan-signmod-1998, oneil-ideas-2007, li-vldb-2014, li-sigmod-2013, goodwin-SIGIR-2017, seshadri-micro-2013, seshadri2017ambit, seshadri-ieeecal-2015, hajinazar-asplos-2021, FastBitA9, wu-icde-1998, guz-ndp-2014,redis-bitmaps,alser-bioinformatics-2017, loving-bioinformatics-2014, xin-bioinformatics-2015, cali2020genasm, kim-genomics-2018, lander-nature-2001, altschul-jmb-1990, myers-jacm-1999})} and are \mkkk{relatively} simple \mk{to support in flash chips}. Data movement is reduced by performing bulk bitwise operations in a massively parallel manner across all flash \mk{cells, \mkf{arrays,} and chips} without moving the data outside flash chips, leading to high performance and \mkv{large} energy savings. \mkkk{Unfortunately}, prior works do \emph{not} investigate \mkv{how to} implement homomorphic operations using IFP. Although HE operations could \mk{potentially} benefit from parallel processing \mk{using} flash \mk{cells, arrays, and chips}, \mkkk{they require \mkn{support for} arithmetic \mkn{operations}, which could be difficult to implement using IFP}.

\subsection{Limitations of Prior HE Accelerators}
\label{sec:accel}
\mayankkk{Existing
HE accelerators adopt large on-chip caches or buffers (180MB for
SHARP~\cite{kim2023sharp}, 256MB for CraterLake~\cite{samardzic2022craterlake}, and 512MB for
BTS~\cite{kim2022bts}) to reduce frequent data loading \mk{from off-chip}.
However, such large on-chip storage may still be insufficient for \mkf{performing} string matching on large encrypted data\mk{bases}. Near-data processing \mkkk{in main memory or SSDs} can be promising since it is challenging \mk{and costly} to \mkf{significantly} increase either the \mkkk{off-chip} memory bandwidth or the on-chip storage.} Prior works \mk{(e.g.,~\cite{gupta2024memfhe,reis2020computing,gupta2023evaluating, zhou2025fhemem, mwaisela2024evaluating}) propose frameworks to perform HE operations near main memory. However, data transfer from SSD to main memory \mk{can} still \mk{limit the performance of} secure string matching \mk{(see \S\ref{sec:pim} \& \S\ref{sec:ndp})}. \mayankkk{Recent works~\cite{yang2023poseidon, suzuki2023designing}}
accelerate arithmetic homomorphic operations by performing computations in \mk{the} SSD \mk{controller}. \mk{However, prior approaches do \emph{not} optimize the data packing schemes, \mkkk{resulting in large memory footprint\mkvvv{s} and inefficient utilization} of \mkvvv{the} overall design (see \S\ref{sec:lim}).} \writing{Moreover,} computations \mk{in} \mkkk{an} SSD controller \mk{is} still limited by the \mkkk{bandwidth of flash memory chips~\cite{park2022flash}} and limited compute capability \mk{and parallelism} due to power constraints} \mk{in the SSD controller~\cite{mailthody-micro-2019, ghiasi2022genstore, kang2013enabling, kim-genomics-2018}.} 
\subsection{Our Goal}
\mkkk{We observe that there is a significant opportunity to exploit in-flash processing (IFP) to perform massively parallel homomorphic encryption-based secure exact string matching using bulk-bitwise computation capabilities of NAND flash memory chips. However, this requires efficient data packing techniques to enable bit-parallel computation and optimization strategies to reduce the complexity of homomorphic operations required for secure string matching.} 

Our goal is to develop an IFP-based hardware-software co-design\mkvvv{ed system that can perform} efficient secure \mknf{exact} string matching. Specifically, we aim to (1) design \mkvv{a memory-}efficient data packing scheme that enables bit-parallel operation and improves performance by eliminating the need for costly homomorphic multiplication operations (\mkvvv{based on Key} Takeaway \ref{sec:take1}), (2) exploit in flash processing to reduce the data movement bottleneck in secure homomorphic string matching (\mkvvv{based on Key} Takeaway \ref{sec:take2}), and (3) overcome the disadvantages of prior methods (Table \ref{tab:compare} and \S\ref{sec:lim}).

%% file: sections/03-overview.tex
\section{CIPHERMATCH: Design } 
\label{sec:ciphermatch}

\subsection{Overview}

\mkv{We propose \textbf{CIPHERMATCH}, an efficient algorithm-hardware co-design\mkvvv{ed system} to accelerate secure string matching. CIPHERMATCH \textbf{algorithm} (see \S\ref{sec:imp1}) comprises two key components: (1) \mkvv{a memory-efficient} data packing scheme to reduce the memory footprint of encrypted data, and (2) a secure string matching algorithm that utilizes only homomorphic addition operations.}
\mkv{CIPHERMATCH \textbf{hardware design} (see \S\ref{sec:ifp}) has two key components: (1) a new in-flash processing (IFP) architecture to accelerate the secure string matching algorithm by reducing the data movement bottleneck, and (2) end-to-end system design to enable CIPHERMATCH in storage (SSD) devices.}

%% file: sections/04-implementation.tex
\subsection{Algorithm Design}
\label{sec:imp1}


\mkv{CIPHERMATCH uses \mkvv{a memory-efficient} data packing scheme to reduce memory footprint and implements a secure string matching algorithm with \mkvv{\emph{only}} homomorphic addition \mkvvv{operations} \mkvv{(i.e. avoiding costly homomorphic multiplications)}. \change{For better understanding, we present the algorithm with specific parameters: \( n = 1024 \), \( q = 32 \), and \( t = 16 \) (see \S \ref{sec:he}). With these parameters, the plaintext polynomial \( P(M) \) has a degree of \( n = 1024 \), with each coefficient of size \( t = 16 \) bits. After encryption, the corresponding plaintext is converted to ciphertext polynomial \( C = (C_0, C_1) \), where both \( C_0 \) and \( C_1 \) have a degree of \( n = 1024 \), and each coefficient is \( q = 32 \) bits in size. Although we explain the algorithm with these specific parameters, the CIPHERMATCH algorithm can be adapted to any set of homomorphic encryption parameters defined by the HE standards~\cite{HomomorphicEncryptionSecurityStandard}.}}

\subsubsection{\mkvv{Memory-Efficient} Data Packing Scheme}
\label{sec:data_packing_mechanism} 
\mkv{Assume a binary string \( P = (b_0, b_1, \dots, b_{k-1}) \) of length \( k \); the string is partitioned into small segments, each containing \( t \) bits, where \( t \) is a predefined (HE) parameter (for our case, \( t = 16 \)). If \( k \) exceeds \( t \), the text is divided into multiple non-overlapping chunks of size \( t \). For example, the first partition, \( T^{(0)}\), will contain the first 16 bits: $ T^{(0)} = (b_0, b_1, \dots, b_{15})$\mkf{,} and the second partition, \( T^{(1)} \), will contain the next 16 bits: $T^{(1)} = (b_{16}, b_{17}, \dots, b_{31})$. We create a packed message  $m_{}(T)$ (see Eqn. (\ref{eq:packing_vector})), which is simply the collection of these partitions:
\begin{equation}
\label{eq:packing_vector}
m_{}(T) = \left( T^{(0)}, T^{(1)}, \dots, T^{\left( \left\lfloor \frac{k}{16} \right\rfloor \right)} \right)
\end{equation}
We convert the packed message \( m(T) \) into a packed plaintext polynomial \( M(x) \) with a maximum degree of \( n \), where \( n \) is a predefined (HE) parameter (in our case, \( n = 1024 \)). The polynomial representation is given by:
$M(x) = \sum_{i=0}^{n-1} m_i x^i$, 
where \( m_i \) represents the packed bits at position \( i \).  
If the number of elements in \( m(T) \) exceeds \( n \), we generate multiple plaintext polynomials. Specifically, if \( m(T) \) contains \( L \) elements, we construct \(\lceil L / n \rceil\) separate polynomials as shown in Eqn. (\ref{eq:packed_vector}):
\begin{equation}
\label{eq:packed_vector}
M^{(j)}(x) = \sum_{i=0}^{n-1} m_{j n + i} x^i, \quad \text{for } j = 0, 1, \dots, \left\lceil \frac{L}{n} \right\rceil - 1.
\end{equation}
We encrypt all the plaintext polynomials with a public key (pk) (see \S \ref{sec:he}) to generate ciphertext $C^{(j)}(x)$ \orphan{(see Eqn. (\ref{eq:encrypted_eqn}))}, with a maximum degree of \(n\) and coefficient size of \(q\), where \(n\), \(q\) are predefined
(HE) parameters (in our case, \( n = 1024 \), \( q = 32 \)): 
\begin{equation}
\label{eq:encrypted_eqn}
\orphan{C^{(j)}(x) = \text{Enc}(M^{(j)}(x), pk), \quad \text{for } j = 0, 1, \dots, \left\lceil \frac{L}{n} \right\rceil - 1.}
\end{equation}
}
\mkn{\textbf{Key Insight.}
Our memory-efficient data packing scheme reduces the memory overhead of encryption, resulting in the encrypted data being only 4$\times$ larger (lower bound - assuming every coefficient of plaintext contains \mknn{the maximum possible packed bits}) than the original unencrypted data (as opposed to 64$\times$ larger (lower bound) in the state-of-the-art data packing scheme~\cite{yasuda2013secure}). \mknn{Our technique achieves this 16$\times$ memory overhead reduction by packing the maximum possible bits (in our case, 16 bits)} into the plaintext coefficients, whereas prior work~\cite{yasuda2013secure} uses a single-bit data packing approach. The 4$\times$ increase in data size after encryption is due to a 2$\times$ increase from converting plaintext into tuples and an additional 2$\times$ increase in ciphertext coefficient size.}

\subsubsection{Secure String Matching Algorithm}
\label{sec:string_search}
\mkv{\mkvvv{We leverage} the \mkvv{memory-efficient} data packing scheme to design an efficient secure string matching algorithm using \mkvvv{\emph{only}} homomorphic addition operations. Assume in plaintext domain, a binary query (\( Q \)); it is first negated ($\sim$\( Q \)) and added to a binary data (\( d \)) such that, if there is a match, the result will be a string of all 1's. This mechanism can be adapted to the HE domain by replacing traditional addition operations with homomorphic addition operations to perform secure string matching. We utilize \mkf{our} \mkvv{memory-efficient} data packing mechanism (\S\ref{sec:data_packing_mechanism}) to efficiently pack the query and input data. 
Let the negated query ($\sim$\( Q \)) and input data (\( d \)) be represented as plaintext polynomials $M_Q(x)$ and $M_d(x)$, respectively, as shown in Eqn. (\ref{eq:poly}):
\begin{equation}
\label{eq:poly}
M_{\sim Q(x)} = \sum_{i=0}^{n-1} \sim Q_i x^i, \quad M_d(x) = \sum_{i=0}^{n-1} d_i x^i
\end{equation}
where $\sim$\( Q_i \) and \( d_i \) are the respective packed binary coefficients (\S\ref{sec:data_packing_mechanism}) of the query (\( Q \)) and the input (\( d \)). The polynomials are then encrypted, and the resulting ciphertexts are denoted as \( C_{\sim Q(x)} \) and \( C_d(x) \), respectively. After encryption, each coefficient of the ciphertext is mapped to a ring  $R_q$, and the 16-bit coefficients are converted into 32-bit coefficients. While the coefficients are transformed during encryption, the original plaintext data stays in the same location; however, it is now represented by its encrypted value at that location.
We perform homomorphic addition on the encrypted polynomials:
$C_{\text{result}} = \textit{Hom-Add}(C_{\sim Q(x)}, C_d(x))$
where \( \textit{Hom-Add} \) represents the homomorphic addition operation.}

\textbf{Index Generation.} \mkv{When a match occurs, the result of the homomorphic addition is a ciphertext where one (or more) coefficients is equal to the encrypted value corresponding to all \( 1 \)'s, i.e., $P_v$(x) = 111...11$x^{1023}$+ ... +111..11. We call this polynomial $P_v$(x) \mkvv{the} "match polynomial". The encrypted result is compared with an encrypted “match polynomial” for each coefficient. This comparison checks whether the values in the encrypted data match those in the "match polynomial". If a match is found, it indicates that the query has successfully found a corresponding value in the input data. Based on this comparison, we determine the exact location where the match occurred. }




We describe how to use this secure string matching algorithm within a client(user)-server model, as shown in Algorithm \ref{tab:algo}.
%
\begin{algorithm}
\vspace{1mm}
\begin{center}
    \textbf{\revision{Database Preparation: (Server Side)}}
\end{center}

\begin{algorithmic}[1]

\State \revision{\textbf{Input: }\texttt{Assume data (d) = (10110001...10) of length \(m>n\).}}

\State \revision{\textbf{Generate: } \texttt{Packed vector \( d = (d_0, \dots, d_{m'}) \), where \( d_i \) are 16-bit partitions, \( m' = \lceil m/16 \rceil \), and construct polynomials \( P_j(d) = \sum_{i=j}^{j+1023} d_i x^i \) for \( j \in (0, m'/1024) \).}}

\State \revision{\textbf{Encrypt:} \texttt{E($P_j(d)$, pk) =  $C_{j}^s(x)$ = ($ct_{0,j}^s$, $ct_{1,j}^s$), store on the server.}}

\noindent\hrulefill

\vspace{-1mm}
\begin{center}
    \textbf{\revision{Query Preparation: (Client/User Side)}}
\end{center}



\State \revision{\textbf{Input:} \texttt{ Assume query (Q) = (001...10) of length $y \leq m$. }}

\State \revision{\textbf{Generate:} \texttt{$Q'= $ $\sim$$Q$ as a packed vector $pv = \left(Q_{0}, \ldots, Q_{\frac{y}{16}}\right)$, where $Q_i$ represents 16 bits.}}

\State \revision{\texttt{If $\frac{y}{16}$<1024, repeat the vector $\left(pv,pv,…,pv\right)$ to fill the coefficients of the plaintext polynomial.}}

\State \revision{\texttt{Construct a polynomial $P\left(Q'\right)$ = $\textstyle{\Sigma}_{i=0}^{1023} Q_i x^i$}}

\State \revision{\texttt{Perform $Q_i' << 1$ for $n$ times and repeat \mk{steps 6-8}.}}

\State \revision{\textbf{Encrypt:} \mk{\texttt{E($P\left(Q’\right)$, pk) =  $C_{i}^u(x)$ = ($ct_{0,i}^u$, $ct_{1,i}^u$). Send this encrypted query to the server.}}}


\noindent\hrulefill

\vspace{-1mm}
\begin{center}
    \textbf{\revision{Secure String Search: (Server Side)}}
\end{center}


\State \revision{\textbf{Homomorphic Addition:} \texttt{ E($P_i\left(Q’\right)$ + $P_j\left(d\right)$) = ($ct_{0,j}^s$ + $ct_{0,i}^u$, $ct_{1,j}^s$ + $ct_{1,i}^u$) for all i and j.}}

\State \revision{\textbf{Output:} \texttt{This step generates multiple resultant ciphertexts ($ct_{0,j}^r$, $ct_{1,j}^r$) between server (s) and user (u) ciphertexts.}}

\State \revision{\textbf{Index Generation:} \texttt{If the ciphertext contains a "match polynomial" in any coefficient, a match is found. The location index is generated and sent back to the client/user.}}

\end{algorithmic}
\caption{\textbf{:} \textbf{CIPHERMATCH algorithm demonstrated in a client(user)-server model.}}
\label{tab:algo}
\end{algorithm}



\noindent\textbf{1) Database Preparation (lines 1-3).} We assume that the database is initially flattened into a binary vector (lines 1-2), which is then transformed into an encrypted packed polynomial (line 3) using our memory-efficient data packing scheme before being stored on the server.

\noindent{\textbf{2) Query Preparation (lines 4-9).}
To efficiently perform secure string matching over a large encrypted database, we employ our memory-efficient data packing scheme that minimizes the memory footprint. We describe query preparation in four key steps that enable efficient parallel string matching.

\squishlist 
\item \textbf{Lines 4-6.} \orphan{Given a client/user query \((Q)\) of length 8 bits (e.g.,  $001\ldots10$) (line 4), we first negate it (line 5) and replicate it multiple times to construct a structured plaintext polynomial (line 6). This ensures that the same query is present across multiple coefficients of the same polynomial. As a result, when homomorphic addition is performed, multiple coefficient-wise additions perform multiple string matching operations (see \S\ref{sec:string_search}) in parallel.} 


\item \textbf{Line 7.} \orphan{The replicated query is embedded within a structured polynomial  $P_1(x)$, where each coefficient stores the same binary pattern:
$P_1(x) = 110\ldots01110\ldots01x^{1023} + 110\ldots01110\ldots01x^{1022} + \dots + 110\ldots01110\ldots01$. Each coefficient fits within \(t = 16\)-bits. If the query length exceeds 16 bits, the bits are distributed across multiple coefficients while ensuring efficient packing.}

\item \textbf{Line 8.} \orphan{To detect all possible alignments of the query within the encrypted database, we generate multiple left-shifted variants of \(Q\). This results in a set of shifted polynomials $P_1(x), P_2(x),\dots, P_8(x)$, where each polynomial represents a different alignment of the query.}

\item \textbf{Line 9.} \orphan{Each shifted polynomial is then homomorphically encrypted using the same HE parameters, producing ciphertext polynomials  $C_1^u(x), C_2^u(x), \dots, C_8^u(x)$, where \(u\) represents the user \orphan{(i.e., client)} query. These encrypted polynomials are then used to find a match against the encrypted database.}
\squishend



\noindent\textbf{3) Secure String Search (lines 10-12).} These encrypted query polynomials are then transferred to the server, which performs an efficient homomorphic addition (\textit{Hom-Add}) (lines 10-11) with the encrypted database. The server generates an index (line 12) by comparing the resultant polynomial with \orphan{the} encrypted "match polynomial".

\subsection{Hardware Design}
\label{sec:ifp}
\orphan{CIPHERMATCH introduces a new in-flash processing architecture (see \S\ref{sec:ifp_architecture}) to perform efficient bitwise operations and enable bit-serial addition operations, using which we can efficiently implement homomorphic addition. To leverage the benefits of this IFP architecture, we discuss the end-to-end system design (see \S\ref{sec:system_intergation}) to integrate CIPHERMATCH in \mkvv{an} SSD.}

\subsubsection{In-Flash Processing (IFP) Architecture}
\label{sec:ifp_architecture}
\mkvvv{Our IFP architecture is inspired by two recent IFP works, ParaBit~\cite{gao2021parabit} and Flash-Cosmos~\cite{park2022flash}. ParaBit enables parallel bitwise operations (AND, OR) by controlling the data and sensing latch circuitry within the flash chip. \change{Flash-Cosmos leverages the existing XOR circuit in NAND flash peripherals to perform bitwise XOR operations and introduces an Enhanced SLC Programming (ESP) method for reliable execution of these operations. Additionally, Flash-Cosmos supports bulk bitwise computation of AND and OR operations across data stored in NAND flash memory. However, there are limitations with both prior \mkf{approaches}. First, ParaBit inherently allows data transfer only from the S-latch to the D-latch and performs bitwise AND and OR operations by controlling one-sided data movement, which restricts the reuse of intermediate or previously computed results. Second, Flash-Cosmos only supports bulk-bitwise operations across the data present in the NAND flash memory.} To perform string matching with a query, it is necessary to write the query \mkn{string} to NAND flash memory \mkn{cells}, which \mkn{adds latency and} accelerates the \mkn{wearout} of the flash memory.}

\mkvvv{Figure~\ref{fig:latch_circuit} illustrates the design of the sensing latch (S-latch) and data latch (D-latch) in NAND flash memory, which supports bitwise (AND, OR, and XOR) operations. In this \mkn{design}, we \mkn{build on} a prior work~\cite{cho-patent-2022}, which adds two transistors (M7 and M8 in \mkf{Figure}~\ref{fig:latch_circuit}) in the existing NAND flash peripheral circuit. We utilize these transistors to enable \emph{bi-directional \mkn{data} flow} (shown as blue arrows in Figure~\ref{fig:latch_circuit}) by controlling the activation of transistors (M1-M8). This enables fine-grained control over bitwise AND and OR operations by explicitly managing the data transfer direction, ensuring that results are stored in the desired latch while allowing efficient reuse of intermediate or previously computed results. We use TLC NAND flash memory to leverage its \mkvv{\emph{multiple data latches}} (see \S\ref{sec:NANDflash}) for intermediate data storage while operating it in SLC mode to ensure reliable computation of bitwise operations using \mkvvv{the} ESP method~\cite{park2022flash}. We explain (i) the data transfer from \mkv{the} S-latch to \mkv{the} D-latch and \mkv{(ii) the} execution of bitwise (AND, OR, and XOR) operations \mkv{using the proposed NAND flash peripheral circuitry (which fundamentally differs from ParaBit~\cite{gao2021parabit}) as follows.}}
\begin{figure}[h]
    \centering
\includegraphics[width=0.8\linewidth]{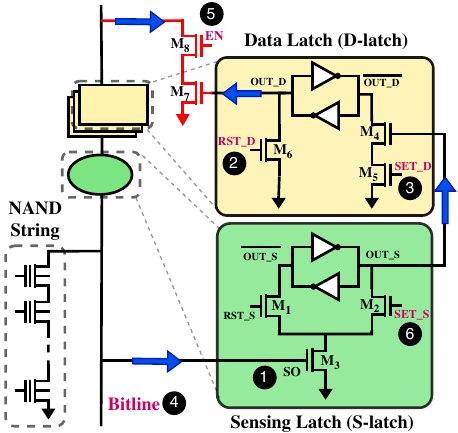}
    \caption{NAND flash peripheral circuitry with S-latch and D-latches to perform bitwise operations. \orphan{Transistors (in red) show the modifications proposed by prior work~\cite{cho-patent-2022}. The arrows (in blue) denote the data flow between the latches.}}
    \label{fig:latch_circuit}
\end{figure}

\noindent\mkv{\textbf{1) Data transfer from S-latch to D-latch} occurs in three key steps. First, data is read from the flash cell to the S-latch~\circledd{1} using the conventional flash read operation (see \S\ref{sec:NANDflash}). Second, the \texttt{M6} transistor is activated (by setting \texttt{RST\_D}~\circledd{2} to logical \texttt{1}) to reset \texttt{OUT\_D} to logical \texttt{0}. Third, enabling the \texttt{M5} transistor (by setting \texttt{SET\_D}~\circledd{3} to logical \texttt{1}) allows \texttt{OUT\_D} to be determined by \texttt{OUT\_S}: if \texttt{OUT\_S} is at logical \texttt{1}, the \texttt{M4} transistor is activated, which pulls $\overline{\texttt{OUT\_D}}$ to \texttt{GND} and set\mkvv{s} \texttt{OUT\_D} to logical \texttt{1}; otherwise, \texttt{OUT\_D} remains at logical \texttt{0}. This is equivalent to copying data from the S-latch to the D-latch.}

\noindent\mkv{\textbf{2) Bitwise AND of operands in S-latch and D-latch} is performed in three key steps. \mk{First, we assume that the \mkvv{input} data is present in the S-latch and the D-latch.} Second, we precharge the bitline~\circledd{4} and enable the \texttt{M8} transistor (by setting \texttt{EN}~\circledd{5} to logical \texttt{1}). If the D-latch holds a logical \texttt{0}, the \texttt{M7} transistor does not turn on, keeping the bitline precharged at logical \texttt{1}; otherwise, it pulls the bitline to \texttt{GND}, representing logical \texttt{0}. Third, we enable \texttt{M2} (by setting \texttt{SET\_S}~\circledd{6} to logical \texttt{1}) and \mk{if the bitline value is logical \texttt{1}}, it pulls \texttt{OUT\_S} to \texttt{GND}, making it logical \texttt{0}, regardless of the previously stored value; otherwise, if the bitline value is logical \texttt{0}, it \mkf{turns off} the \texttt{M3} transistor, and \texttt{OUT\_S} retains its previously stored value. This is equivalent to performing an AND operation between the values present in the D-latch and the S-latch and storing the result in \emph{only} the S-latch.} 

\noindent\mkv{\textbf{3) Bitwise OR of operands in S-latch and D-latch} is similar to performing data transfer from S-latch to D-latch. Instead of resetting the \texttt{OUT\_D} value, the \texttt{OUT\_S} value from the S-latch is transferred to the D-latch. We enable \texttt{M5} (by setting \texttt{SET\_D}~\circledd{3} to logical \texttt{1}) of the target D-latch. If \texttt{OUT\_S} is logical \texttt{1}, this pulls $\overline{\texttt{OUT\_D}}$ to \texttt{GND}, setting \texttt{OUT\_D} to \texttt{1}. If \texttt{OUT\_S} is logical \texttt{0}, $\overline{\texttt{OUT\_D}}$ and \texttt{OUT\_D} retain their current values. This is equivalent to performing an OR operation between the values present in the D-latch and the S-latch and storing the result in \emph{only} the D-latch.}

\noindent\textbf{4) Bitwise XOR of operands in D-latches} is possible to perform using the existing peripheral circuit in most modern NAND flash chips~\cite{kim-ieeejssc-2018, park2022flash, chen2024search}. An \texttt{XOR} circuit is present in between two D-latches to perform on-chip data randomization during write operations \cite{kim-ieeejssc-2012} or built-in error detection during chip testing \cite{cao-elecletters-2022}. In our design, we use this XOR circuitry to perform an \texttt{XOR} operation between D-latches 1 and 2, storing the result in D-latch 1. \mk{Flash-Cosmos~\cite{park2022flash} utilizes this \texttt{XOR} circuitry to perform Boolean operations.}

CIPHERMATCH \mkv{uses the proposed IFP architecture} to enable the bit-serial addition mechanism to perform homomorphic addition across multiple bitlines in a flash array, processing data in parallel within each flash chip. We describe the required data layout and the process for performing bit-serial addition.


\textbf{Data Layout.} \mkv{To perform bit-serial addition for 32-bit operands (see \S\ref{sec:data_packing_mechanism}), the key challenge is the \mkf{carry propagation} across all bit positions. In conventional NAND flash memory, operands are stored in a horizontal data layout, where the bits of each operand are placed contiguously in the cells of one wordline. When these bits are read, they are stored in the sensing latches connected to each bitline. This layout restricts the carry propagation because the carry bit must be moved between sensing latches as we perform the addition.
To address this issue, we adopt a vertical data layout, where the bits of the operands are arranged along bitlines instead of wordlines~\cite{hajinazar-asplos-2021}. In this layout, the bits of each operand are distributed across multiple wordlines, with each bit of a 32-bit element stored in a separate bitline. This enables us to compute the carry bit for each position while keeping it stored in a D-latch for the next bit position’s calculation, facilitating bit-serial addition~\cite{batcher1982bit}.}

\textbf{Bit-Serial Addition.} 
In the bit-serial addition process, a full-adder circuit is required to compute the sum and carry for each bit. The full-adder takes two operand bits, \texttt{A\textsubscript{i}} and \texttt{B\textsubscript{i}}, and an input carry, \texttt{C\textsubscript{i}}, and produces a sum bit \texttt{S\textsubscript{i}} \texttt{=} \texttt{A\textsubscript{i}} $\oplus$ \texttt{B\textsubscript{i}} $\oplus$ \texttt{C\textsubscript{i}} and a carry-out bit \texttt{C\textsubscript{o}} \texttt{= (}\texttt{A\textsubscript{i}} $\oplus$ \texttt{C\textsubscript{i}}) $\cdot$ \texttt{B\textsubscript{i}} + \texttt{A\textsubscript{i}} $\cdot$ \texttt{C\textsubscript{i}}. The carry must then propagate through all bit positions as the addition proceeds. Figure \ref{fig:addition} shows the sequence of operations for adding one bit of an input value \texttt{B} transferred to the SSD, to one bit of the value \texttt{A} stored in the flash memory. The carry-in value is initially stored in D-latch 2 and set to 0. 
~\circled{1} Load the input bit $B_i$ from the controller to the S-latch, and
~\circled{2} copy it to D-latch 1.
~\circled{3} Perform the \texttt{AND} operation to calculate \texttt{B\textsubscript{i}} $\cdot$ \texttt{C\textsubscript{i}} in \manos{the} S-latch and
~\circled{4} perform the \texttt{XOR} operation to calculate \texttt{B\textsubscript{i}} $\oplus$ \texttt{C\textsubscript{i}} in D-latch 1.
~\circled{5} Copy the result of the \texttt{AND} operation to D-latch 0.
~\circled{6} Read the value \texttt{A\textsubscript{i}} from the NAND flash cell and 
~\circled{7} copy it to D-latch 2. 
~\circled{8} Transfer the data from D-latch 1 to S-latch to perform  \texttt{A\textsubscript{i}}$\cdot$(\texttt{B\textsubscript{i}} $\oplus$ \texttt{C\textsubscript{i}}). 
~\circled{9} Perform the \texttt{XOR} operation between D-latch 1 and 2 to calculate \texttt{A\textsubscript{i}} $\oplus$ \texttt{B\textsubscript{i}} $\oplus$ \texttt{C\textsubscript{i}}.~\circledd{10} Copy \texttt{A\textsubscript{i}}$\cdot$(\texttt{B\textsubscript{i}} $\oplus$ \texttt{C\textsubscript{i}})   to D-latch 2 and
~\circledd{11} copy \texttt{B\textsubscript{i}} $\cdot$ \texttt{C\textsubscript{i}} to the S-latch.~\circledd{12} \manos{Perform the \texttt{OR}} operation in D-latch 2 \manos{to calculate (\texttt{B\textsubscript{i}} $\oplus$ \texttt{C\textsubscript{i}}) $\cdot$ \texttt{A\textsubscript{i}} + \texttt{B\textsubscript{i}} $\cdot$ \texttt{C\textsubscript{i}} as carry bit}. ~\circledd{13} The sum bit \texttt{S\textsubscript{i}}, which is stored in D-latch 1, is sent out to the SSD controller and the carry-out bit \texttt{C\textsubscript{o}} is stored in D-latch 2. CIPHERMATCH calculates the next sum bit by repeating this procedure to \mkf{calculate the sum of input B \mknf{and} input A}.

\begin{figure*}[h]
    \centering
    \includegraphics[width=0.9\linewidth]{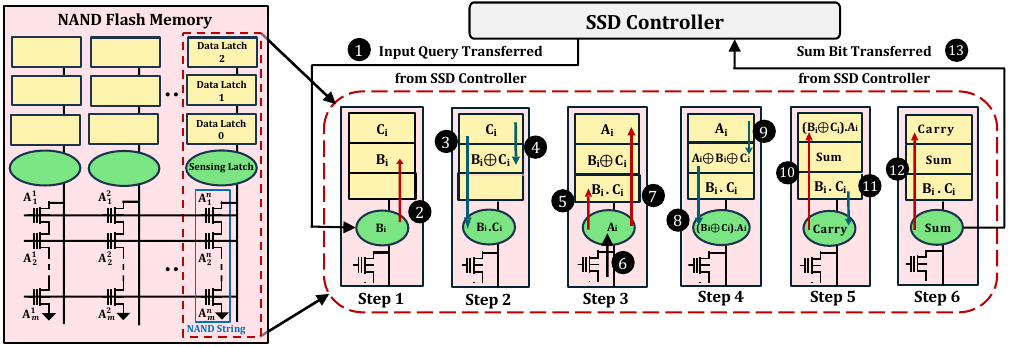}
    \caption{A single step of bit-serial addition using sensing and data latch circuitry in the flash array. Red, blue, and black lines denote data transfer from S-latch to D-latch, bitwise operations, and flash read operations, respectively. }
    \label{fig:addition}
\end{figure*}

\mkvvv{\textbf{Implementing Homomorphic Addition.} 
In homomorphic encryption, ciphertexts (encrypted data) are represented as polynomials, with each coefficient stored as a large integer (in our case, 32 bits). Homomorphic addition performs an element-wise summation of these coefficients, which can be performed independently and in parallel. Our NAND flash architecture supports bit-serial addition, enabling efficient execution of homomorphic addition directly in flash memory. We organize the 32-bit coefficient of a polynomial along a single bitline and distribute polynomial coefficients across multiple bitlines. Bit-serial addition is performed within each bitline by serially summing the 32 bits of \mkn{the} stored coefficient with the coefficient of the query. To maximize throughput, we exploit multiple levels of parallelism: bitline-level parallelism across wordlines, chip-level parallelism across NAND chips, and channel-level parallelism for concurrent coefficient-wise addition operation.}



\textbf{Reliability.} \change{CIPHERMATCH maintains the reliability of bitwise operations and the underlying flash memory using two key strategies. First, CIPHERMATCH leverages enhanced SLC mode programming~\cite{park2022flash} to maximize the voltage difference between the two programming states within a cell, ensuring accurate data representation during bit-serial operations on individual bits. Second, CIPHERMATCH performs bit-serial addition completely using the latching circuit present in the flash chips, which avoids performing costly program/erase (P/E) cycle operations in the flash cell array that degrade the lifetime of flash memory.}

\subsubsection{End-to-End System Design}
\label{sec:ete-system-support}
\label{sec:systemsupport}
\change{In this section, we present the end-to-end system design changes required to enable CIPHERMATCH in storage (i.e., in an SSD). We describe a system-level overview of the CIPHERMATCH algorithm, followed by a detailed explanation of the design of the individual components \change{and system-level modifications to} integrate CIPHERMATCH in an SSD.}

\textbf{System-level Overview.}
\mkv{Figure~\ref{fig:overview} shows a system-level overview of \mayank{CIPHERMATCH.} We assume the database is first packed and encrypted with the optimized packing scheme (see \S\ref{sec:data_packing_mechanism}) before being stored on the server. CIPHERMATCH follows a six-step procedure to perform secure string matching.}
~\circled{1} The client(user) machine prepares the encrypted query, and the encrypted "match polynomial", and ~\circled{2} sends them to the server. 
\circled{3}
\harshita{\mayankkk{The server forwards them to \mkf{the} SSD controller \mkv{(using our proposed system interface)}}} 
and triggers a $\mu$-program for homomorphic addition, consisting of a sequence of flash commands (see \S\ref{sec:ifp_architecture}).~\circled{4} 
\harshita{The $\mu$-program executes} homomorphic addition and utilizes array-level and bit-level parallelism of flash memory.~\circled{5} Finally, an index generation step (see \S\ref{sec:string_search}) to identify the matching location is performed in the
SSD controller and~\circled{6} \harshita{ the index is sent back to the client(user).}

\begin{figure}[h]
    \centering
    \includegraphics[width=\linewidth]{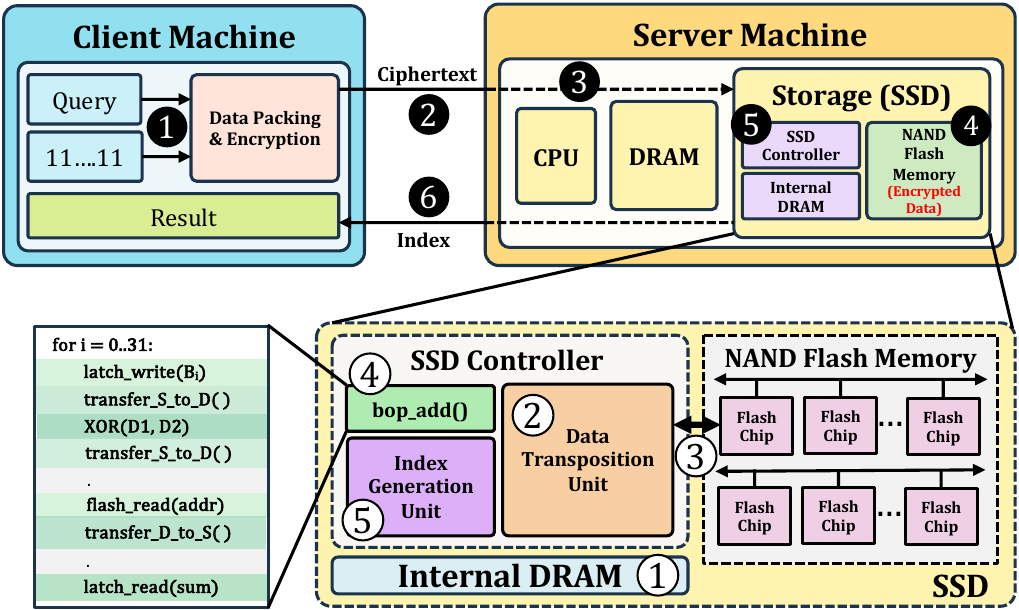}
    \caption{System-level overview of CIPHERMATCH.}
    \label{fig:overview}
\end{figure}


\textbf{System Integration.}
\label{sec:system_intergation}
We explain the necessary system changes to leverage CIPHERMATCH within the SSD from the host device.

\noindent\revision{\textbf{1) Operations in Vertical Data Layout.} The CIPHERMATCH algorithm relies on bit-serial addition, which requires the data to be stored vertically \change{(we store the data across 32 wordlines \mk{to perform 32-bit addition}). However, CPUs traditionally work with data stored in a horizontal layout.} To address this challenge, CIPHERMATCH partitions \mkv{the physical address space of an SSD} into two regions: the conventional storage region and the CIPHERMATCH storage region. We handle the reads/writes for both regions separately. The conventional storage region handles read/write operations in TLC mode with a horizontal layout, while the CIPHERMATCH region operates in SLC mode with a vertical layout.}

\revision{We explain the read and write operations in the CIPHERMATCH region as follows. For writes, CIPHERMATCH (i) encrypts the data, (ii) stores it in the DRAM buffer~\circleddcolor{1}, (iii) performs data transposition in the SSD controller~\circleddcolor{2} \mk{(see Figure ~\ref{fig:overview})} to prepare the data in a vertical layout, and (iv) \mkf{write} the vertical data to the flash chip~\circleddcolor{3}. Each region maintains a separate mapping table for logical to physical address mapping, ensuring that data transposition remains transparent to the programmer. \mk{We introduce a software-based data transposition unit for this purpose (see Figure \ref{fig:overview})}. The address mapping for the vertical layout is managed within its designated mapping table. Separate mapping tables handle commands for the conventional and CIPHERMATCH regions. For reads, the SSD controller reads NAND flash pages sequentially and stores them in the DRAM buffer. \mk{This data is transposed back into a horizontal layout using the software-based data transposition in the SSD controller before sending it back to the host.}} Prior works~\cite{wong2024tcam,chen2024search} \mk{have} explored \mk{such} firmware modifications, and we provide an overhead analysis in \S \ref{sec:eval}.

\vspace{1em}
\noindent\revision{\textbf{2) Handling Page Faults to the CIPHERMATCH Region.}} 
\mkn{A page fault occurs when an application accesses data that is not present in DRAM, requiring the host to issue a read request to the SSD. In the CIPHERMATCH region, data is stored in the vertical layout, where each 32-bit coefficient is distributed across multiple bitlines (see \S\ref{sec:ifp_architecture}). If the page fault happens to the CIPHERMATCH region, the SSD controller performs flash reads from multiple wordlines to read the homomorphically encrypted data and transpose\mknn{s} the data to the conventional horizontal layout (using the data transposition unit). This read request experiences a longer latency due to reading \emph{multiple} (e.g., 32) flash wordlines; however, we do not account for the latency for data transposition as it can be performed in parallel with flash reads (see \S\ref{sec:firmware_modification}). }
\mkn{To handle these long-latency page faults efficiently, CIPHERMATCH leverages the operating system (OS) support for huge pages~\cite{ganapathy1998general}. \mknn{The page fault handler in the OS maintains a timeout (a configurable threshold) for these long latency read requests that defines the maximum wait time before retrying the request to prevent long delays.}}


\noindent\orphan{\textbf{3) Handling Dirty Writebacks to the CIPHERMATCH Region.} A dirty writeback occurs when an application evicts modified data from the main memory (dirty data) that has not been written to the SSD. The operating system (OS) issues a write request to the SSD, where data is asynchronously written to the NAND flash memory. If the write request happens in the CIPHERMATCH region, the SSD controller transposes the data (using the data transposition unit) and writes it to NAND flash memory asynchronously at a granularity of 4KB (page size). CIPHERMATCH leverages the SSD’s existing write policies to manage these writes. Since these writes are asynchronous, data transposition has a minimal impact on the overall performance of the running application.}


\noindent\revision{\textbf{4) System Interfaces.} 
\change{We add three new commands to control the CIPHERMATCH region so that applications can leverage the benefits of CIPHERMATCH.} We add a new flag to the conventional I/O read and write commands and design new interfaces, \texttt{CM-read and CM-write}, to handle data in the vertical format. When these new commands are used, the flag informs the SSD controller \mkf{to transpose data} while reading or writing the data. This 1-bit flag differentiates between conventional operations and the CIPHERMATCH region, activating the transposition unit as required. For search operations using CIPHERMATCH, we design a new command \texttt{CM-search}, which contains the encrypted query as a parameter. The host device transmits the \texttt{CM-search} command and encrypted query, which is then transferred to the data latches to initiate the bit serial addition (see \S \ref{sec:ifp_architecture}).}


\textbf{Firmware Modifications.}
\label{sec:firmware_modification}
\mk{We} explain the modification\mk{s to} \mkf{the} SSD firmware to \mkv{enable} CIPHERMATCH operations.

\noindent\revision{\textbf{1) \(\mu\)-Program for Addition.} CIPHERMATCH utilizes three key operations to perform bit-serial addition (see \S \ref{sec:ifp_architecture}): (i) data transfer from S-latch to D-latch, (ii) Bitwise (XOR, AND and OR) operations between S and D latch\mkvv{es}, (iii) read operation to transfer the data from the flash cell to \mkf{the} S-latch. To \mknf{efficiently} enable \mknf{the} bit-serial addition operation, \mknf{we introduce} a new command, \texttt{bop\_add} (bulk operation add)~\circleddcolor{4} \mkv{in the flash translation layer (FTL)}. (see Figure \ref{fig:overview}). This command executes a sequence of operations that performs bit-serial addition across multiple wordlines within NAND flash chips (see Figure \ref{fig:addition}), leveraging the inherent parallelism of NAND flash memory.}
\texttt{CM-search} command calls the \texttt{bop\_add} command to perform the search operation.

\noindent\revision{\textbf{2) Data Transposition Unit.} CIPHERMATCH uses a software-based data transposition unit~\circleddcolor{2} (see Figure ~\ref{fig:overview}) running on \mkf{the} SSD controller to handle \texttt{CM-read} and \texttt{CM-write} operations and page faults. The transposition unit operates at a 4KB granularity, corresponding to the NAND flash page size. \mkvv{We use a software-based data transposition unit instead of a hardware unit (see \S\ref{sec:transposition}) because the latency of software-based data transposition latency using the SSD controller is 13.6$\mu$s, which is lower than flash read latency~\cite{park2022flash, cheong2018flash} and can be efficiently overlapped with flash read operations, minimizing the overall performance impact.}}

\noindent\revision{\textbf{3) Index Generation Unit.}  The final index generation of the matched location is performed by the SSD controller~\circleddcolor{5} as part of the \texttt{CM-search} command. The result of the homomorphic addition is transferred from the D-latches to the SSD controller, which compares it with the encrypted “match polynomial” to generate the index of the matched location (see \S \ref{sec:string_search}). We \mkf{estimate} the total latency of the index generation step running on \mkf{the} SSD controller cores (see \S\ref{sec:met}) to be 3.42$\mu$s, which can be
effectively overlapped with the sequential flash read latency.}

%% file: sections/05-methodology.tex
\section{Methodology}
\label{sec:met}



\orphan{Our evaluation methodology consists of three key components: (1)~a real system evaluation methodology to assess the performance improvement provided by the software implementation of the CIPHERMATCH algorithm on an existing real CPU system (see \S\ref{sec:sw_analysis}), (2)~a simulation-based evaluation methodology to analyze the benefits of implementing the CIPHERMATCH algorithm using IFP (see \S\ref{sec:hw_analysis}), and (3)~details of evaluated workloads (see \S\ref{sec:workloads}).}


\subsection{Real System Evaluation Methodology}
\label{sec:sw_analysis}

\noindent\textbf{Real System Configuration.} Table \ref{tab:config1} presents the configuration of the real CPU system used in our experiments. The system comprises \mkf{of} six out-of-order CPU cores, 32 GB of DDR4-2400 main memory, and a PCIe 4.0 NVMe SSD. We conduct our experiments to \mkvvv{measure} the increase in memory footprint and the performance of homomorphic operations. 


\begin{table}[ht]
   \centering
   \footnotesize
   \renewcommand{\arraystretch}{1.3}
   \resizebox{\columnwidth}{!}{
   \begin{tabular}{c|l}
   \hline
   \multirow{5}{*}{\centering \shortstack{\textbf{CPU:} \\ \textbf{Intel(R) Xeon(R)} \\ \textbf{Gold 5118 }} } 
   & \textit{Microarchitecture:} Intel Skylake~\cite{intelskylake} \\ \cline{2-2} 
   & x86-64~\cite{guide2016intel}, 6~cores, out-of-order, 3.2 GHz  \\ \cline{2-2}
   & \emph{L1 Data + Inst. Private Cache:} 32kB, 8-way, 64B line \\ \cline{2-2}
   & \emph{L2 Private Cache:} 256kB, 4-way, 64B line \\ \cline{2-2}
   & \emph{L3 Shared Cache:} 8MB, 16-way, 64B line \\
   \hline
   \textbf{Main Memory} & 32GB DDR4-2400, 4~channels \\
   \hline
   \textbf{Storage (SSD)} & Samsung 980 Pro PCIe 4.0 NVMe SSD 2 TB~\cite{samsung-980pro} \\
   \hline
   \textbf{Operating System (OS)} & Ubuntu 22.04.1 LTS\\
   \hline
   \end{tabular}
   }
   \caption{Real CPU system configuration.}
   \label{tab:config1}
\end{table}

\noindent\textbf{Evaluated Systems.} We evaluate the software-based CIPHERMATCH algorithm (CM-SW) implemented using Microsoft SEAL~\cite{sealcrypto} and compare its performance and energy consumption against two state-of-the-art approaches: (i) \mkf{the} Boolean approach~\cite{aziz2024secure} with TFHE-rs~\cite{TFHE-rs} and (ii) \mkf{the} arithmetic approach~\cite{yasuda2013secure} with Microsoft SEAL~\cite{sealcrypto}.

\noindent\textbf{Performance and Energy Measurements.} We measure the execution time of our baselines and CM-SW using performance counters to focus specifically on the time spent executing homomorphic operations, excluding the overhead associated with encryption and encoding. To estimate energy consumption, we combine power parameters obtained from Intel’s RAPL tool~\cite{desrochers2016validation, khan2018rapl} with the execution time obtained from our performance measurements.
\subsection{Simulation-Based Evaluation Methodology} 
\label{sec:hw_analysis}

\textbf{\mkvvv{Simulated} Systems.} We \mkvvv{simulate} four different systems, including CIPHERMATCH implemented on our IFP architecture (CM-IFP) \mkvvv{using our in-house simulator} to understand the trade-offs between compute-centric, memory-centric, and storage-centric processing approaches: 


\squishlist
\item \textbf{CM-SW (compute-centric)} is a conventional CPU implementation of the CIPHERMATCH algorithm. The goal is to identify the data movement bottlenecks and SIMD limitations in existing compute-centric systems.


\item \textbf{CM-PuM (memory-centric)} is a processing-using memory approach in which we implement the CIPHERMATCH algorithm on an external 32GB DDR4 DRAM (see Table \ref{tab:config2}) and use \mkvvv{the} SIMDRAM framework~\cite{hajinazar-asplos-2021} to perform \mkvvv{string matching} computations. \change{The data is transferred from the SSD to the external DRAM, where parallel bit-serial addition operations are performed across multiple channels and banks. Using this bit-serial addition, homomorphic addition is executed to implement the CIPHERMATCH algorithm. The goal is to evaluate performance improvements by reducing data movement between the DRAM and the CPU while leveraging \mkf{the parallelism of DRAM} for computation. }

\item \textbf{CM-PuM-SSD (storage-centric)} is a processing-in storage approach in which we implement the CIPHERMATCH algorithm within the SSD’s 2GB LPDDR4 DRAM, following the SIMDRAM semantics (similar to CM-PuM). \change{It utilizes the internal flash channel bandwidth for efficient data transfer between NAND flash memory and \mkvvv{SSD-internal} DRAM. The goal is to evaluate performance improvements by reducing data movement through external I/O while utilizing the parallelism of \mkvvv{SSD-internal} DRAM.}
\squishend

\begin{table}[ht]
    \centering
    \footnotesize
    \renewcommand{\arraystretch}{1.3}
    \resizebox{\columnwidth}{!}{
    \begin{tabular}{c|l}
    \hline
    \multirow{6}{*}{\centering \textbf{CM-PuM}} & 32 GB DDR4-2400, 4 channel, 1 rank, 16 banks; \\
    & Peak throughput: 19.2 GB/s \\ \cline{2-2} 
    & \textbf{Latency:} \texttt{T\textsubscript{bbop}}: 49 ns;  \textbf{Energy:} \texttt{E\textsubscript{bbop}}: 0.864 nJ; \\ 
    & where $bbop$ is bulk bitwise operation \\ \cline{2-2} 
    & \textbf{SSD External-Bandwidth}: 7-GB/s external I/O bandwidth; \\ &(4-lane PCIe Gen4) \\ 
    \hline
    \multirow{14}{*}{\centering \shortstack{\textbf{CM-IFP} \\ \textbf{and} \\  \textbf{CM-PuM-SSD}}} & 48-WL-layer 3D TLC NAND flash-based SSD; 2 TB\\ \cline{2-2} 
    & \textbf{SSD Internal DRAM}: 2GB LPDDR4-1866 DRAM cache;\\
    & 1 channel, 1 rank, 8 banks\\ \cline{2-2} 
    & \textbf{\mkvvv{NAND-Flash Channel Bandwidth}}: 1.2-GB/s Channel IO rate\\ \cline{2-2} 
    & \textbf{Controller Cores}: ARM Cortex-R5 series @1.5GHz; 5 Cores~\cite{arm-cortexR5} \\ \cline{2-2} 
    & \textbf{NAND Config}: 8 channels; 8 dies/channel; 2 planes/die;\\
    & 2,048 blocks/plane; 196 (4$\times$48) WLs/block; 4 KiB/page\\ \cline{2-2} 
    & \textbf{Latency}: \texttt{T\textsubscript{read}} (SLC mode): 22.5 $\mu$s~\cite{park2022flash}; \texttt{T\textsubscript{AND/OR}}: 20 ns~\cite{gao2021parabit}; \\  
    & \texttt{T\textsubscript{latchtransfer}}: 20 ns~\cite{gao2021parabit}; \texttt{T\textsubscript{XOR}}: 30 ns~\cite{park2022flash}; \texttt{T\textsubscript{DMA}}: 3.3 $\mu$s;\\ 
    & \mkvvv{\texttt{T\textsubscript{bit\_add}} (CM\_IFP): 29.38$\mu$s}\\ \cline{2-2} 
    & \textbf{Energy}: \texttt{E\textsubscript{read}} (SLC mode): 20.5$\mu$J/channel ~\cite{park2022flash}; \\ &
    \texttt{E\textsubscript{AND/OR}}: 10nJ/KB~\cite{gao2021parabit}; \texttt{E\textsubscript{latchtransfer}}: 10nJ/KB~\cite{gao2021parabit}; 
    \\ & \texttt{E\textsubscript{XOR}}: 20nJ/KB~\cite{park2022flash}; \texttt{E\textsubscript{DMA}}: 7.656$\mu$J/channel; \\ & \texttt{E\textsubscript{index\_gen}} (SSD controller): 0.18$\mu$J/page size;\\ 
    & \texttt{E\textsubscript{bit\_add}} \texttt{(CM\_IFP)}: 32.22$\mu$J/channel\\ 
    \hline
   \end{tabular}
   }
  \caption{\mkvvv{Simulated} system configurations.}
   \label{tab:config2}
\end{table}


\noindent\textbf{Performance Modeling.} Table \ref{tab:config2} presents different \mkvvv{simulated} near-data processing (NDP) system configurations. We develop an in-house simulator to model the SSD and external DRAM latency characteristics accurately. We model three different NDP systems and compare \mkf{them} against a simulated CM-SW\footnote{For a fair comparison, we model the CPU-based system \textbf{(CM-SW)} by incorporating CPU computation, DRAM transfer, SSD read, and I/O transfer latency, based on the CPU configuration outlined in Table \ref{tab:config1}.} baseline.
\squishlist

\item \textbf{For CM-PuM}, we develop a SIMDRAM-based simulator that models the latency of in-DRAM bitwise XOR, AND, and OR operations for implementing bit-serial addition. We use these values along with the DRAM configuration (see Table \ref{tab:config2}) in the simulator to compute the latency of bit-serial addition. Additionally, we model external I/O bandwidth and DRAM bandwidth to model data movement \mkvvv{and its overheads} accurately.

\item \textbf{For CM-PuM-SSD}, \mkvvv{we compute the latency of bit-serial addition using the same SIMDRAM-based simulator. We use the SSD-internal DRAM configuration to compute the latency of bit-serial addition. Additionally, we model the NAND-flash channel bandwidth and SSD-internal DRAM bandwidth to model overall data movement and its overheads.}

\item \textbf{For CM-IFP}, we develop our custom simulator and model three key operations required to enable CIPHERMATCH (see \S\ref{sec:ifp_architecture}): (i) bitwise XOR, AND, and OR operations, (ii) latch transfer latency, and (iii) NAND flash read latency (see Table \ref{tab:config2}). We input the NAND flash configuration to our simulator to calculate the latency of our bit-serial addition algorithm (see Table \ref{tab:config2}). For SSD controller tasks, such as data transposition and index generation (see \S\ref{sec:ete-system-support}), we develop a C program and run it in a QEMU environment~\cite{qemu} configured for the ARM Cortex-R5 series~\cite{arm-cortexR5} to obtain the latency parameters. \orphan{We calculate the latency of one-bit serial addition (\texttt{T\textsubscript{bit\_add}}) using Eqn.~\eqref{eq:latency_bitserial}: }
\begin{equation}
\label{eq:latency_bitserial}
\orphan{T_{bit\_add} = T_{bop\_add} + 2*T_{DMA}}
\end{equation}
\begin{equation}
\label{eq:latency_bop}
\orphan{T_{bop\_add} = T_{read} + 2*T_{XOR} + 5*T_{latchtransfer} + 4*T_{AND/OR}}
\end{equation}
\squishend

\noindent\textbf{Energy Modeling.}
To model DRAM
energy consumption, we use the power values based on a
DDR4 \mkvvv{power} model \cite{ddr4sheet, ghose2019demystifying}.  
To model SSD energy consumption, we use the SSD power values of Samsung 980 Pro SSDs~\cite{samsung-980pro} and the NAND flash power values measured by Flash-Cosmos~\cite{park2022flash}. Using \mkf{these} power values, we \mkf{estimate} energy consumption of DRAM, SSD, and NAND flash \mkvvv{using} our performance models. \change{To calculate the energy consumption of CM-IFP, we use an equation similar to Eqn.~\eqref{eq:latency_bop}, replacing the time variable with the energy consumption of bulk operations.} Additionally, we include the energy consumed by the SSD controller during the index generation phase (\texttt{E\textsubscript{index\_gen}}) (see Table~\ref{tab:config2}). \orphan{We calculate the energy consumption of one-bit serial addition (\texttt{E\textsubscript{bit\_add}}) using Eqn.~\eqref{eq:energy_bitserial}:}
\begin{equation}
\label{eq:energy_bitserial}
\orphan{E_{bit\_add} = E_{bop\_add} + 2*E_{DMA} + E_{index\_gen}}
\end{equation}

\subsection{Workloads} 
\label{sec:workloads} 
We evaluate \harshita{the performance of CIPHERMATCH using two real-world workloads:}
1) exact DNA string matching and 2) encrypted database search. 

\textit{1) Exact DNA String Matching:} 
DNA sequence analysis relies on exact string matching in the seeding operation to extract smaller reads/bases from a reference genome~\cite{cali2020genasm,firtina2023rawhash,bhukya2011exact,de2013secure, alser2020accelerating, alser2022molecules,aziz2024secure}. The query sizes for these reads \mkvvv{can} range from 8 to 128 base pairs. For performance evaluation, we use a single query and vary the query sizes between 16 bits and 256 bits. We design a synthetic workload that uses a 32GB DNA database, which grows to 128GB (larger than the external DRAM size) once encrypted with our \mkvvv{memory-efficient data packing scheme (see \S\ref{sec:data_packing_mechanism}).}

\textit{2) Encrypted Database Search:} In scenarios where a client needs to search for specific records on a server while preserving privacy and security, secure string matching is performed on an encrypted database~\cite{silberschatz1997database,raj2020web,kim2021efficient, koudas2004flexible}. For performance evaluation, we design a synthetic workload that varies the database size from 2GB to 32GB, which grows from 8GB to 128GB upon encryption with our optimized packing scheme. We simulate 1000 queries to assess how different database sizes affect performance under two conditions: when the entire encrypted database fits in DRAM and when it must be fetched repeatedly from the SSD.


%% file: sections/06-evalutation.tex
\section{Evaluation}
\label{sec:evaluation}




\orphan{We evaluate the effectiveness of CIPHERMATCH \mkn{at} improving the performance and energy consumption of secure string matching. \S\ref{sec:cm_sw_analysis} evaluate\mkn{s the} performance and energy consumption of software-based CIPHERMATCH (CM-SW). \S\ref{sec:cm_ifp_analysis} evaluates \mkn{the} performance and energy consumption of hardware-based CIPHERMATCH  (CM-PuM, CM-PuM-SSD, and CM-IFP). \S\ref{sec:eval} studies the overheads of implementing CM-IFP on off-the-shelf SSDs.}

\subsection{Software-Based CIPHERMATCH Analysis}
\label{sec:cm_sw_analysis}

\subsubsection{Effect of Query Size} 
\begin{figure}[h]
    \centering
        \centering
        \includegraphics[width=\linewidth]{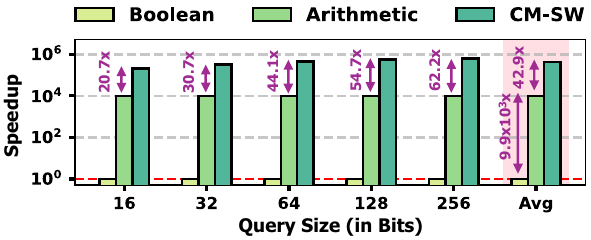}
        \caption{\orphan{Speedup (higher is better) of the arithmetic approach~\cite{yasuda2013secure} and CM-SW over the Boolean approach~\cite{aziz2024secure} (denoted as Y=\texttt{10\textsuperscript{0}}) for different query sizes, with a database size of 128GB and a single query.}}
        \label{fig:software_genome}
\end{figure}

\mkvvv{Figure~\ref{fig:software_genome} presents the speedup of the arithmetic approach~\cite{yasuda2013secure} and CM-SW over the Boolean approach~\cite{aziz2024secure} (denoted as Y=\texttt{10\textsuperscript{0}}) for different query sizes ranging from 16 bits to 256 bits. For this analysis\mkf{,} we use a single query \mkn{and a database size of 128GB} (see \S\ref{sec:workloads}).}

We make \change{three} key observations. \mkvvv{First,} CM-SW outperforms the arithmetic~\cite{yasuda2013secure} and Boolean~\cite{aziz2024secure} approaches by \mkvvv{20.7-62.2$\times$} and \mkvvv{2.0$\times10^5$ - 6.2$\times10^5$$\times$} for different query sizes, respectively. 
\mkvvv{Second,} CM-SW speedup over the arithmetic approach~\cite{yasuda2013secure} \mkn{increases} with query size. \change{For example, CM-SW speedup} increases from 20.7$\times$ to 30.7$\times$ when we increase the query size from 16 bits to 32 bits. 
This improvement is because CM-SW uses \emph{only} homomorphic addition for secure string matching, which is computationally \mkn{much} less expensive than the arithmetic approach~\cite{yasuda2013secure}, \mkf{which} uses \mkn{costly} arithmetic homomorphic operations \mkn{(e.g., multiplication) and more homomorphic operations are needed with an increase in query size.} 
\orphan{Third, our evaluation shows that CM-SW consistently outperforms prior works across all query sizes.}

\begin{figure}[h]
    \centering
        \centering
        \includegraphics[width=\linewidth]{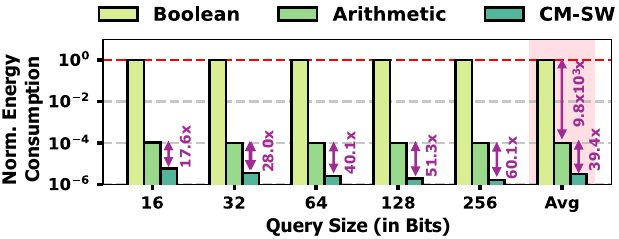}
        \caption{\orphan{Energy consumption (lower is better) of the arithmetic approach~\cite{yasuda2013secure} and CM-SW normalized to the Boolean approach~\cite{aziz2024secure} (denoted as Y=\texttt{10\textsuperscript{0}}) for different query sizes, with a database size of 128GB and a single query.}}
        \label{fig:software_genome_energy}
\end{figure}

\mkvvv{Figure~\ref{fig:software_genome_energy} presents the energy consumption of the arithmetic approach~\cite{yasuda2013secure} and CM-SW normalized to the Boolean approach~\cite{aziz2024secure} for different query sizes ranging from 16 bits to 256 bits. For this analysis, we use a single query \mkn{and a database size of 128GB} (\S\ref{sec:workloads}).}

We make \change{three} key observations. \mkvvv{First,} \mkvvv{CM-SW reduces the energy consumption by 17.6-60.1$\times$ and 1.6$\times10^5$ - 6.0$\times10^5$$\times$ compared to the arithmetic~\cite{yasuda2013secure} and Boolean~\cite{aziz2024secure} approaches for different query sizes, respectively.} 
\mkvvv{Second, the energy consumption \mkn{benefits} of CM-SW over the arithmetic approach~\cite{yasuda2013secure} \mkn{increase} with query size. For example, CM-SW reduces energy consumption from 17.6$\times$ to 28.0$\times$ over the arithmetic approach~\cite{yasuda2013secure} when we increase the query size from 16 bits to 32 bits.}
\orphan{Third, our evaluation shows that CM-SW consistently provides energy savings over prior works across all query sizes.}



\subsubsection{Effect of Database Size}
\begin{figure}[h]
    \centering
        \includegraphics[width=\linewidth]{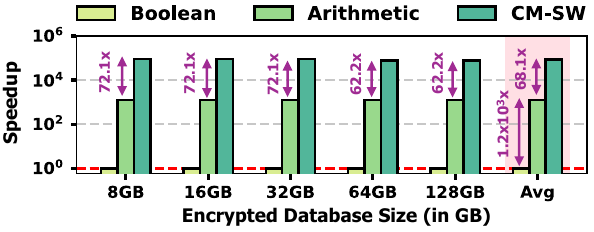}
        \caption{\orphan{Speedup (higher is better) of the arithmetic approach~\cite{yasuda2013secure} and CM-SW over the Boolean approach~\cite{aziz2024secure} (denoted as Y=\texttt{10\textsuperscript{0}}) for different encrypted database sizes, with a query size of 16 bits and 1000 queries.}}
        \label{fig:software_database}
\end{figure}

\mkvvv{Figure~\ref{fig:software_database} presents the speedup of the arithmetic approach~\cite{yasuda2013secure} and CM-SW over the Boolean approach~\cite{aziz2024secure} for different encrypted database sizes ranging from 8GB to 128GB. For this analysis, we use 1000 queries with a query size of 16 bits (see \S\ref{sec:workloads}). }


\vspace{1em}
We make \change{three} key observations. \mkvvv{First,} CM-SW outperforms the arithmetic~\cite{yasuda2013secure} and Boolean~\cite{aziz2024secure} approaches by \mkvvv{62.2-72.1$\times$ and 7.6$\times$$10^4$ - 8.8$\times$$10^4$$\times$} for different encrypted database sizes, respectively. \mkvvv{Second,} when the database size exceeds 32GB, the performance \mkn{improvement} of CM-SW reduces by 1.16$\times$. This is because the data movement overhead in CM-SW increases due to frequent I/O transfers for each query. In contrast, both the arithmetic~\cite{yasuda2013secure} and Boolean~\cite{aziz2024secure} approaches saturate the memory footprint beyond the external DRAM size, even for smaller database sizes (e.g., 8 GB), due to their different data packing schemes. 
\orphan{Third, our evaluation shows that CM-SW consistently outperforms prior works across all database sizes.}

\noindent \mknf{\textbf{Summary.}} \mknn{We conclude that CM-SW \mknnn{provides} significant performance improvement and energy savings over prior state-of-the-art approaches~\cite{yasuda2013secure,aziz2024secure} across various query and database sizes. This improvement is primarily due to two reasons: (i) the use of \emph{only} \mknnn{the} homomorphic addition operation that is computationally less expensive than homomorphic multiplication and (ii) the use of a memory-efficient data packing scheme that reduces both memory footprint and data movement.  }

\subsection{Hardware-Based CIPHERMATCH Analysis}
\label{sec:cm_ifp_analysis}


\subsubsection{Effect of Query Size} \label{sec:impact_of_query_sizes}
\begin{figure}[h]
    \centering
        \includegraphics[width=\linewidth]{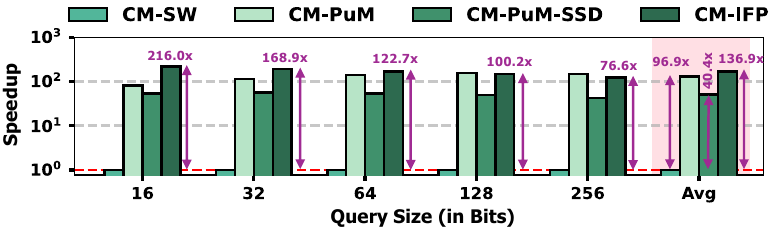}
        \caption{\orphan{Speedup (higher is better) of CM-PuM, CM-PuM-SSD, and CM-IFP over CM-SW (denoted as Y=\texttt{10\textsuperscript{0}}) for different query sizes, with a database size of 128GB and a single query.}}
        \label{fig:hardware_genome}
\end{figure}

\mkvvv{Figure~\ref{fig:hardware_genome} presents the speedup of CM-PuM, CM-PuM-SSD, and CM-IFP over CM-SW (denoted as Y=\texttt{10\textsuperscript{0}}) for different query sizes ranging from 16 bits to 256 bits. For this analysis, we use a single query \mkn{and a database size of 128GB} (\S\ref{sec:workloads}).} 

We make \change{five} key observations. \mkvvv{First,} CM-IFP, CM-PuM-SSD, and CM-PuM outperform CM-SW by \mkvvv{76.6-216.0$\times$, 81.7-105.8$\times$, and 26.4-53.9$\times$,} for different query sizes, respectively. These improvements are due to \mkn{two reasons: (i) reduced data movement to the compute units and (ii) improved parallelism at the array-level and bit-level to perform homomorphic additions.} \mkvvv{Second,} CM-IFP outperforms CM-PuM-SSD by \mkvvv{2.89-4.03$\times$} for different query sizes. This improvement is due to two reasons: (i) data is not transferred outside the flash memory \mkn{chips}, reducing the data movement latency, and (ii) CM-IFP utilizes flash array, chip, and channel-level parallelism, which is higher than that of the SSD-internal DRAM to achieve higher throughput for computations. 

Third, for small query sizes (e.g., 16-bit), CM-IFP outperforms CM-PuM by 2.64$\times$; however, for large query sizes (e.g., 256-bit), CM-PuM achieves 1.21$\times$ higher performance than CM-IFP. We observe that CM-IFP speedup over CM-SW varies with query size. For example, CM-IFP speedup reduces from 216.0$\times$ to 168.9$\times$ over CM-SW when we increase the query size from 16 bits to 32 bits. This is because for one query, even after reducing the data movement, performing computations for a large query size (e.g., 256 bits) using DRAM reads are significantly faster than flash reads. Fourth, CM-PuM outperforms CM-PuM-SSD by \mkvvv{1.5-3.5$\times$} for different query sizes. While CM-PuM-SSD reduces data movement by performing computations within the SSD and utilizing NAND-flash channel bandwidth, its performance is limited by the smaller internal DRAM, restricting the throughput of homomorphic additions compared to external DRAM. 
\orphan{Fifth, our evaluation shows that CM-IFP provides the highest average performance across all query sizes.}

\begin{figure}[h]
    \centering
        \centering
        \includegraphics[width=\linewidth]{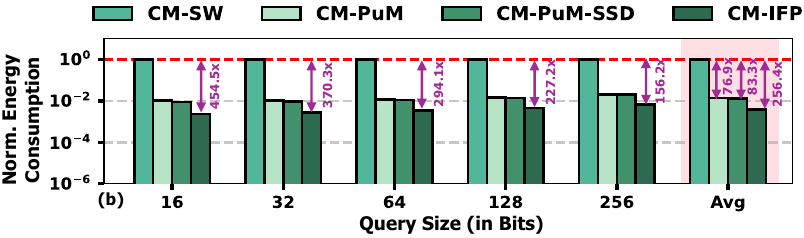}
        \caption{\orphan{Energy consumption (lower is better) of CM-PuM, CM-PuM-SSD, and CM-IFP normalized to CM-SW (denoted as Y=\texttt{10\textsuperscript{0}}) for different query sizes, with a database size of 128GB and a single query.}}
        \label{fig:hardware_genome_energy}
\end{figure}

\mkvvv{Figure~\ref{fig:hardware_genome_energy} presents the energy consumption of CM-PuM, CM-PuM-SSD, and CM-IFP normalized to CM-SW for different query sizes ranging from 16 bits to 256 bits. For this
analysis, we use a single query \mkn{and a database size of 128GB} (see \S\ref{sec:workloads}).} 

We make \change{three} key observations. First, CM-IFP, CM-PuM-SSD, and CM-PuM reduce the energy consumption compared to CM-SW by \mkvvv{156.2-454.5$\times$, 49.1-111.8$\times$, and 48.6-98.3$\times$} for different query sizes, respectively. Second, while CM-PuM outperforms CM-PuM-SSD in terms of performance \mkn{(see Figure ~\ref{fig:hardware_genome})}, CM-PuM-SSD is 1.06$\times$ more energy efficient on average than CM-PuM. This is due to the lower energy cost of data transfers through internal NAND flash channels compared to external I/O channels.
\orphan{Third, our evaluation shows that CM-IFP provides the highest average energy savings across all query sizes.}

\subsubsection{Effect of Database Size} 
\begin{figure}[h]    
        \centering
        \includegraphics[width=\linewidth]{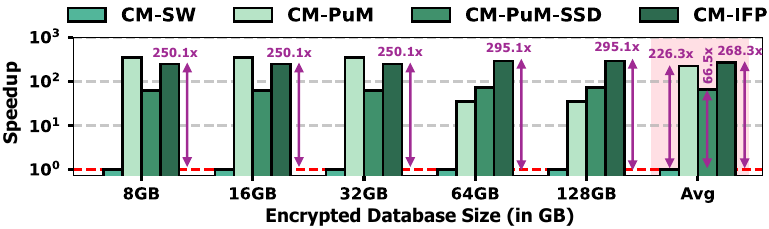}
        \caption{\orphan{Speedup (higher is better) of CM-PuM, CM-PuM-SSD, and CM-IFP over CM-SW (denoted as Y= \texttt{10\textsuperscript{0}}) for different encrypted database sizes, with a query size of 16 bits and 1000 queries.}}
        \label{fig:hardware_database}
    
\end{figure}

\mkvvv{Figure~\ref{fig:hardware_database} presents the speedup of CM-PuM, CM-PuM-SSD, and CM-IFP over CM-SW for different encrypted database sizes ranging from 8GB to 128GB. For this analysis, we use 1000 queries with a query size of 16 bits (see \S\ref{sec:workloads}). }

We make \change{four} key observations. \mkn{First, CM-IFP, CM-PuM-SSD, and CM-PuM outperform CM-SW by 250.1-295.1$\times$, 52.8-62.3$\times$, and 35.6-353.4$\times$ for different query sizes, respectively.} Second, we compare CM-IFP and CM-PuM to understand the impact of SSD-external bandwidth. For small encrypted databases ($\leq$ 32GB, which fits in external DRAM), CM-PuM outperforms CM-IFP by 1.41$\times$; however, for large encrypted databases ($>$ 32GB, exceeding \mkn{the size of} external DRAM), CM-IFP achieves $8.29\times$ higher performance than CM-PuM. This is because, for smaller database sizes, the data can be loaded \mkn{and then mostly accessed from} DRAM for multiple queries and \mkn{CM-PuM amortizes the cost of} external I/O transfers \mkn{to the SSD}. However, for a large database that exceeds the DRAM capacity, the data must be fetched from the SSD, leading to frequent data movement \mkn{between DRAM and SSD} that degrades performance \mkn{with CM-PuM}.

Third, we compare CM-PuM and CM-PuM-SSD to understand the impact of NAND-flash channel bandwidth. \mkn{For small encrypted databases ($\leq$ 32GB, \mkn{which fits} in external DRAM), CM-PuM outperforms CM-PuM-SSD by 6.6$\times$; however, for large encrypted databases ($>$ 32GB, exceeding \mkn{the size of} external DRAM), CM-PuM-SSD achieves 1.75$\times$ higher performance than CM-PuM. This is because, for smaller database sizes, the limited capacity of internal DRAM restricts the throughput of CM-PuM-SSD in homomorphic addition operations. In contrast, CM-PuM leverages extensive channel and bank-level parallelism. However, for a large database and multiple queries, CM-PuM-SSD utilizes NAND-flash channel bandwidth, reducing data movement time and outperforming CM-PuM, which requires frequent external I/O transfers \mkn{to the SSD}.} 
\orphan{Fourth, our evaluation shows that CM-IFP provides the highest average performance across all database sizes.}

\noindent \mknf{\textbf{Summary.}} \mknn{We draw two major conclusions from our analysis of hardware-based CIPHERMATCH implementation\mknnn{s}. \change{First, \mknnn{all three} hardware-accelerated CIPHERMATCH implementations (CM-PuM, CM-PuM-SSD, and CM-IFP) provide significant performance improvements and energy savings over the pure software-based CIPHERMATCH implementation (CM-SW) across various query and database sizes. This improvement is primarily due to (i) reduced data movement from/to main memory and/or storage devices and (ii) a significant increase in throughput by utilizing array-level and bit-level parallelism to perform homomorphic addition operations. Second, the IFP implementation of CIPHERMATCH (CM-IFP) provides the highest performance improvements and energy savings across all query and database sizes.}}

\subsection{Overhead Analysis of CM-IFP}
\label{sec:eval}


\revision{\textbf{Storage Overhead.} CM-IFP incurs three types of storage overhead. First, CIPHERMATCH partitions the SSD into two regions, i.e., SLC mode for the CIPHERMATCH region and TLC mode for the conventional region (see \S \ref{sec:ete-system-support}), reducing the overall capacity of the SSD. Second, the IFP bit serial algorithm sends the results to the controller for \mkn{index generation (see \S\ref{sec:ete-system-support})}, which requires additional storage in the \mkn{SSD-}internal DRAM. Based on our SSD configuration (see Table \ref{tab:config2}), CIPHERMATCH occupies 0.5 MB of \mkn{SSD-internal} DRAM space \emph{(4KB (page size) $\times$ 8 (number of channels) $\times$ 8 (number of dies) $\times$ 2 (number of planes))} for storing \mkn{the results of homomorphic addition}. Third, CIPHERMATCH stores the $\mu$-program ($bop\_add$) in the internal-SSD DRAM, which requires less than 1KB.}


\noindent \textbf{Area Overhead.} \mkn{CM-IFP can be performed with minimal hardware modification to the NAND flash memory. CIPHERMATCH is built on two key designs: 1) hardware modifications proposed by ParaBit~\cite{gao2021parabit} to enable bitwise (AND and OR) operations, which incurs 0.6\% area overhead to modify the NAND-flash peripheral circuitry, and 2) Flash-Cosmos's ESP technique~\cite{park2022flash} (see \S\ref{sec:ifp_architecture}) to enable reliable bitwise operations, which does \emph{not} incur any area overhead. Hence, the area overhead is $\sim$0.6\% of the die area of the baseline NAND flash memory.} 

%% file: sections/07-discussion.tex
\section{Discussion}

\subsection{Hardware\mkn{-Based Data} Transposition Unit}
\label{sec:transposition}


CIPHERMATCH uses a software-based data transposition unit to perform data transposition using the SSD’s general-purpose core. Data transposition latency can be effectively overlapped with flash read latency (22.5$\mu$s) by pipelining the transposition operation with multiple flash reads.
However, with advancements in flash memory technology that further reduces the read latency (e.g., 3$\mu$s in the latest Z-NAND flash memory~\cite{cheong2018flash}), we need to accelerate data transposition to hide the transposition latency \orphan{during flash reads}. To this end, we propose the integration of a dedicated hardware transposition unit adjacent to the SSD controller. The design of this unit is similar to SIMDRAM’s data transposition unit~\cite{hajinazar-asplos-2021}. We implement the hardware-based data transposition unit in Verilog HDL and synthesize our design using a 22nm CMOS technology node~\cite{gf22} to estimate the latency and area overhead\mknnn{s}. The data transposition unit performs the transposition of 4KB data in 158ns and incurs an area overhead of 0.24 $mm^{2}$.

\subsection{Mitigation Techniques for Privacy Concerns}
\label{sec:privacy_concerns}
CIPHERMATCH returns the index of the matched location to the user after performing secure string matching (see \S\ref{sec:string_search}), and this index must be securely transmitted from the server to the user to ensure privacy.
To securely transfer the index of the matched location, we leverage a hardware-based 256-bit AES module present in commodity SSDs~\cite{reidys2022rssd, samsung-ssd} to encrypt the index before transmission across vulnerable channels (e.g., from an SSD in the cloud to the client system). 
During an offline step, the SSD controller generates a new AES key and sends the key to the client system. To secure the AES key during transfer, the SSD controller encrypts it with a public-key encryption algorithm before sending it to the client, mitigating the risk of exposure over vulnerable channels. This encryption occurs in an offline step, allowing the cost of key transfer to be amortized.
We encrypt the index with the AES key and send the encrypted index back to the user/client.
We implement the AES hardware unit (which works at 16 bytes granularity) using Verilog HDL and synthesize it using 22nm CMOS technology node~\cite{gf22}. The latency to encrypt 16 bytes is estimated at 12.6ns, with an area overhead of 0.13 $mm^{2}$.

%% file: sections/08-related-work.tex
\section{Related Work}

\mkv{To our knowledge, CIPHERMATCH is the first in-flash processing system designed to accelerate secure} \mknf{exact} string matching. \mknnn{We have already qualitatively and quantitatively compared both software and hardware implementations of CIPHERMATCH to the state-of-the-art Boolean~\cite{aziz2024secure,pradel2021privacy} and arithmetic~\cite{yasuda2013secure,kim2017private,bonte2020homomorphic} approaches (see \S\ref{sec:lim} \& \S\ref{sec:evaluation}). In this section, we discuss other related works.}

\textbf{Secure String Matching.} Prior works \mknnn{(e.g.,~\cite{faber2015rich,hahn2018practical,mainardi2019privacy,chase2015substring,guo2024gridse,moataz2013boolean})} have explored \mknnnn{symmetric searchable encryption (SSE)}~\cite{poh2017searchable} for secure string matching, leveraging conventional encryption primitives~\cite{daemen1999aes,liestyowati2020public}.
\mkn{Faber et al.~\cite{faber2015rich} \mknnnn{\mkf{improve} SSE-based secure string matching by using} an optimized data structure like partition trees~\cite{chan2010optimal} to represent the database and use conventional encryption techniques to perform secure string matching.}
\mkn{Hahn et al.~\cite{hahn2018practical} enable efficient and secure substring searches over encrypted data outsourced to untrusted servers by using frequency-hiding order-preserving encryption (OPE)~\cite{kerschbaum2015frequency} techniques.} Mainardi et al.~\cite{mainardi2019privacy} \mkn{improve the secure string matching algorithm to achieve a sub-linear (polylogarithmic) data transfer (communication) complexity between the client and the server.}
\mknnn{Unfortunately}, \mknnnn{SSE-based} schemes exhibit substantial leakage profiles, making them vulnerable to plaintext recovery attacks~\cite{grubbs2017leakage}. 
\mknnn{In contrast}, homomorphic encryption (HE)~\cite{ogburn2013homomorphic} is based on \mknnn{strong} mathematical foundations and has \mknnn{been} proven to be secure~\cite{HomomorphicEncryptionSecurityStandard}.
CIPHERMATCH leverages homomorphic encryption for secure string matching \mknnn{and develops new software and hardware techniques to reduce HE's} computational and memory overheads.

\mkv{\textbf{Near-Data Processing for HE.} Prior works on memory-centric computing \mknnn{(e.g.,~\cite{gupta2024memfhe,zhou2025fhemem,gupta2023evaluating,reis2020computing,chen2024hmc})}
\mkn{focus on performing homomorphic operations within main memory to} reduce data movement \mknnn{overheads}. However, these works \mknnn{can} still \mknnn{be} limited by \mknnn{data movement from/to storage devices, especially} when \mkn{the} dataset \mknnnn{size} exceeds the \mkn{size of} main memory. Other works utilize FPGAs (e.g.,~\cite{yang2023poseidon,suzuki2023designing}), \mkn{to perform homomorphic operations near storage, reducing the data movement from storage \mknnn{devices} to \mknnn{the} CPU}. \mkn{These solutions are} constrained by limited computational capability and NAND-flash channel bandwidth, resulting in high latency and energy consumption for large datasets. 
\mkn{CIPHERMATCH addresses these challenges by performing homomorphic addition operations \emph{directly within} the NAND flash memory \mknnn{using operational principles of flash memory circuitry}, \mknnn{thereby} \mknnnn{both} reducing data movement and \mknnn{efficiently exploiting} large-scale array-level and bit-level parallelism.}}


%% file: sections/09-conclusion.tex
\section{Conclusion}


\mkv{We \mkn{introduce} CIPHERMATCH, \mkn{an algorithm-hardware co-designed system} for accelerating secure \mknf{exact} string matching in solid-state drives \mkn{(SSDs)}. CIPHERMATCH consists of two key components: (1) \mknnn{a memory-efficient data packing scheme that reduces the memory footprint after encryption and eliminates the use of costly homomorphic \mknnnn{operations (i.e.,} multiplication), and}
(2) \mknnnn{an in-flash processing (IFP) architecture that exploits array-level and bit-level parallelism in NAND flash memory to enable parallel string matching through homomorphic computations, requiring only small modifications to NAND flash chips in commodity SSDs.}
\mknnn{Our evaluation shows \mknnnn{that} both pure software-based and hardware-accelerated CIPHERMATCH implementations provide \mknnnn{large} performance improvements and energy savings over state-of-the-art approaches \mknnnn{for homomorphic encryption-based secure \mknf{exact} string matching}.}}

\section*{Acknowledgments}
We thank the anonymous reviewers of ASPLOS 2025 for feedback. We thank the SAFARI group members for feedback and the stimulating intellectual environment they provide. We acknowledge the generous gifts from our industrial partners, including Google, Huawei, Intel, and Microsoft. This work is supported in part by the ETH Future Computing Laboratory (EFCL), Huawei ZRC Storage Team, Semiconductor Research Corporation, AI Chip Center for Emerging Smart Systems (ACCESS), sponsored by InnoHK funding, Hong Kong SAR, and European Union’s Horizon programme for research and innovation [101047160 - BioPIM].